\documentclass[aps,prb,twocolumn,amsmath,amssymb,preprintnumbers,floatfix]{revtex4}

\usepackage[utf8]{inputenc}
\usepackage{microtype}
\usepackage{txfonts}
\usepackage{eucal}
\usepackage{dsfont}
\usepackage{bm} 

\usepackage{enumerate}
\usepackage{color}

\usepackage{graphicx}
\usepackage{float}

\usepackage[colorlinks,allcolors=blue]{hyperref}
\usepackage[capitalize]{cleveref}

\let\mathbb=\mathds

\renewcommand{\d}[1]{\mathrm{d}{#1}}                            

\newcommand{\sgn}{\mathrm{sgn}}                                 

\renewcommand{\i}{\mathrm{i}}

\newcommand{\up}{\uparrow}                                      
\newcommand{\dn}{\downarrow}                                    

\newcommand{\ie}{\textit{i.e.}\ }
\newcommand{\eg}{\textit{e.g.}\ }
\newcommand{\etal}{\textit{et al.}\ }

\definecolor{Jacob}{rgb}{0.65,0,0}
\definecolor{Ali}{rgb}{0.55,0,0.65}
\definecolor{Gray}{rgb}{0.60,0.60,0.60}

\begin{document}
\title{Spin-switch Josephson junctions with magnetically tunable $\textbf{sin}\bm{(\delta\varphi/n)}$ {shape}}
\author{Jabir Ali Ouassou}
\author{Jacob Linder}
\affiliation{QuSpin Center of Excellence, Department of Physics, Norwegian University of Science and Technology, N-7491 Trondheim, Norway}
\date{\today}
 
\begin{abstract}
  \noindent
  With a combination of simple analytical arguments and extensive numerical simulations, we theoretically propose a Josephson junction with $n+1$ superconductors where the current-phase relation can be toggled \emph{in situ} between a $\sin(\delta\varphi)$ and $\sin(\delta\varphi/n)$ shape using an applied magnetic field.
  Focusing in particular on the case $n=2$, we show that by using realistic system parameters such as unequal interface transparencies, the $\sin(\delta\varphi/2)${-shaped solution} retains its $2\pi$-periodicity {due to discontinuities} at $\delta\varphi = \pm \pi$.
  Moreover, we demonstrate that as one toggles between the $\sin(\delta\varphi)${-} and $\sin(\delta\varphi/2)${-shaped solutions}, the system acts as an on--off switch, and can acheive more than two orders of magnitude difference between the supercurrent in the on and off states.
  Finally, we argue that the same approach can be generalized to switchable $\sin(\delta\varphi/n)$ junctions for arbitrary integers~$n${, which we motivate by analytically solving the Josephson equations for double- and triple-barrier junctions.}\\
\end{abstract}

\maketitle

\begin{figure}[t]
  \centering
  \vspace{-2ex}
  \includegraphics[width=\columnwidth]{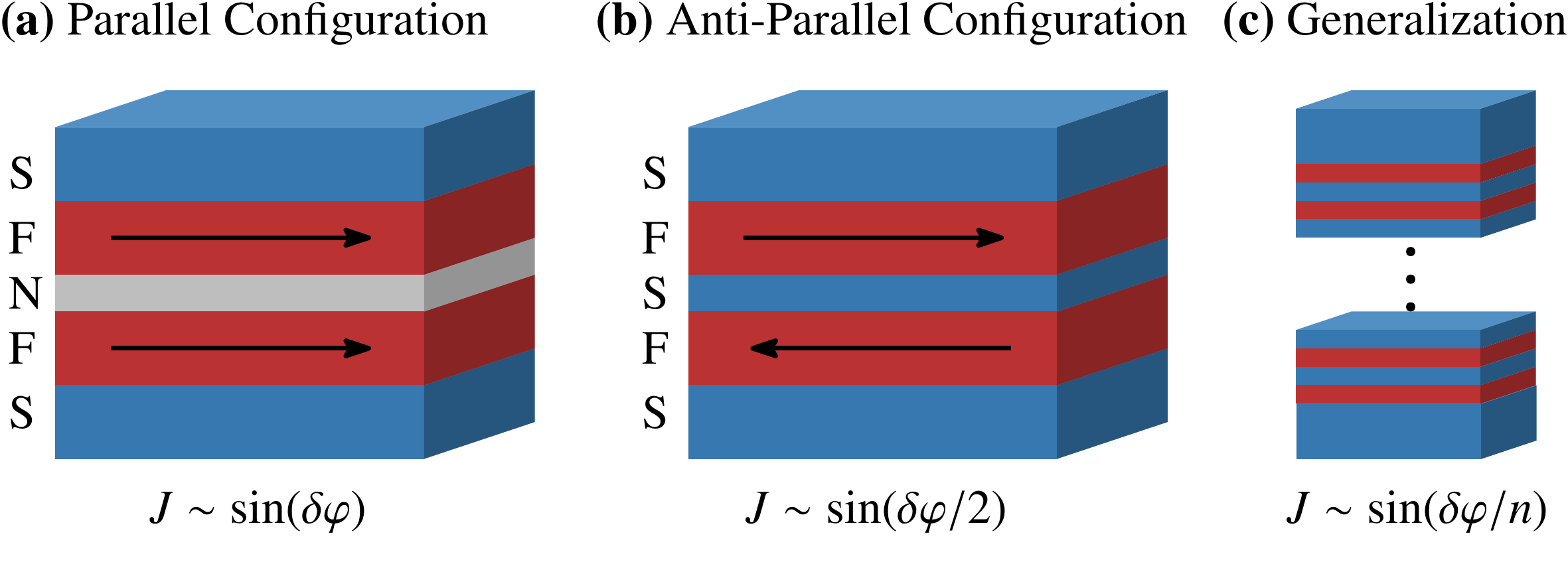}
  \vspace{-5ex}
  \caption
  {%
    Schematic of the proposed device. 
    \textbf{(a)}
    If the ferromagnets are aligned in parallel, the effective exchange fields experienced by the conduction electrons add up inside the superconductor, resulting in a strong net field there.
    This net exchange field suppresses superconductivity, making the central layer act as a normal metal.
    We therefore get an effective S/X/S junction with a conventional $\sin(\delta\varphi)$ current-phase relation.
    \textbf{(b)}
    If the ferromagnets have an antiparallel alignment, their fields cancel near the center of the superconductor.
    The central layer thus acts as a superconductor, and we get an S/X/S/X/S junction with a more exotic $\sin(\delta\varphi/2)$ shape for the current-phase relation instead.
    \textbf{(c)}
    In principle, the same idea can be extended to a junction that consists of $m$ superconductors sandwiched inbetween ferromagnets.
    This should result in a junction with a general $\sin(\delta\varphi/n)$ shape for the current-phase relation, where $n\in[1,\,m+1]$ is magnetically tunable.
    For instance, an S/X/S/X/S/X/S system could support $n=1$ for an $\up\up\up$ configuration of the magnetic layers, $n=2$ for $\up\up\dn$, and $n=3$~for~$\up\dn\up$.
  }%
  \label{fig:model}
\end{figure}

\section{Introduction}\noindent
In a Josephson junction, two superconductors are proximity-coupled through a non-superconducting region, such as a normal metal or insulator.
In conventional Josephson junctions, it can be shown that the current flowing between the superconductors in the tunneling limit is proportional to $\sin(\delta\varphi)$, where $\delta\varphi$ is the phase-difference between the superconductors~\cite{Golubov2004}.
These are also known as 0-junctions, as the ground state is~$\delta\varphi=0$.
In recent years, there has been a lot of interest in developing Josephson junctions with unconventional current-phase relations.
One of the first was the experimental realization of $\pi$-junctions using magnetic elements between the superconductors~\cite{Buzdin2005,Ryazanov2001}, where the current is proportional to $\sin(\delta\varphi+\pi)$, \ie it flows in the opposite direction to a $0$-junction for the same phase-difference~$\delta\varphi$.
{This work was then extended to $\varphi$-junctions with two ground states $\delta\varphi=\pm\varphi$ by combining 0- and $\pi$-junctions~\cite{Goldobin2010,Goldobin2012}.}
Another important development was the very recent construction \cite{Szombati2016} of a $\varphi_0$-junction using spin-orbit interactions~\cite{Buzdin2008,reynoso_prl_08, zazunov_prl_09, tanaka_prl_09, tanaka_prb_10}, where the current was found to be proportional to $\sin(\delta\varphi+\varphi_0)$, with an electrically controllable phase-bias $\varphi_0$.

In this paper, we focus on a different scenario, namely a $\sin(\delta\varphi/n)$ junction.
The special case of $\sin(\delta\varphi/2)$ has previously been discussed in Refs.~\onlinecite{Zapata1996,Ishikawa2001,Kupriyanov1999,DeLuca2009}, and has recently been subject of rekindled interest as a possible signature for a Majorana-mediated supercurrent \cite{beenakker_rcmp_13, kitaev_pu_01, zhang_prb_14, wiedenmann_natcom_16}.
Here, we demonstrate theoretically a Josephson junction where the current-phase relation can be toggled \textit{in situ} between a $\sin(\delta\varphi)$ and $\sin(\delta\varphi/2)$ shape by changing the configuration of a spin-valve via a magnetic field.
The current-phase relation nevertheless retains its $2\pi$-periodicity {due to discontinuities} at $\delta\varphi=\pm\pi$. Moreover, we show that when toggling between the two current-phase relations, the system acts as an on--off switch: the supercurrent magnitude can differ by more than two orders of magnitude in the two states.
We further argue that the same approach can be used to construct more general junctions with $\sin(\delta\varphi/n)$ shapes, where $n$ is an arbitrary and magnetically tunable integer.
In addition to being interesting from a fundamental physics point of view, discovering novel kinds of Josephson junctions may also find applications in both digital and quantum computing \cite{Feofanov2010}.


\section{Analytical argument}
\subsection{{Double-barrier junction}}\noindent
From here on, we use the notation S for superconductors and X for non-superconductors, where X can be any combination of ferromagnets (F), insulators (I), and normal metals (N). 
It is well-known that {in the tunneling limit, the current-phase relation for a single-barrier S/X/S Josephson} junction is \cite{Golubov2004} 
\begin{align}
  J = J_c \sin(\delta\varphi),
\end{align}
where $J_c$ is the critical charge current, and $\delta\varphi$ is the net phase-difference across the junction.
Under certain conditions, this result can be generalized to junctions with multiple superconducting elements. 
Let us first assume that the superconducting phase changes slowly as function of position inside each superconducting layer,
\ie that the supercurrent is {relatively} small.
In this limit, we can treat an S/X/S/X/S junction (see \cref{fig:model}) as a concatenation of two S/X/S subjunctions, where the subjunctions are described by the current-phase relations
\begin{align}\label{eq:currents}
  J_{12} = J_{c} \sin(\varphi_2-\varphi_1), \;\;\;\; J_{23} = J_{c} \sin(\varphi_3-\varphi_2), 
\end{align}
where $J_{ij}$ is the current from superconductor number $i$ to $j$, $\varphi_i$~is the phase of superconductor number $i$, and we have assumed that the critical current of each subjunction is equal.
In any real physical system, these critical currents will not be identical, and we later discuss in detail how this influences the result.
Note that we have some freedom when choosing these phases~$\varphi_i$, since only \emph{phase-differences} affect the physics of the system.
We may therefore define $\varphi_1 \equiv -\delta\varphi/2$ and $\varphi_3 \equiv +\delta\varphi/2$, such that the net phase-difference across the junction is $\varphi_3-\varphi_1=\delta\varphi$:
\begin{align}\label{eq:varphi2-eq}
  J_{12} = J_{c} \sin(\delta\varphi/2+\varphi_2), \;\;\;\; J_{23} = J_{c} \sin(\delta\varphi/2-\varphi_2).
\end{align}
Since the current has to be conserved through the junction, we have the constraint $J_{12} = J_{23} \equiv J$.
The resulting equations has two distinct solutions $\varphi_2=0$ and $\varphi_2=\pi$, yielding the currents
\begin{align}
  J = \pm J_{c} \sin(\delta\varphi/2).
\end{align}
Adding up the energies $E_{ij} = {E_c}[1 - \cos(\varphi_j - \varphi_i)]$ of each subjunction, {where $E_c = \hbar J_c/2e$ is the Josephson energy,} we also find the corresponding junction energies
\begin{align}
  E = {E_c}[{2} \mp {2}\cos(\delta\varphi/2)].
\end{align}
Similarly, one can show that an S/X/S/X/S/X/S junction {results in a $\sin(\delta\varphi/3)$ current-phase relation shape [see \cref{sec:sinphi3}], and that adding more superconductors and barriers in this way may lead to {more general} $\sin(\delta\varphi/n)$ shapes [see \cref{sec:sinphin}]}.


\begin{figure}[b] 
  \centering
  \begingroup
    \makeatletter
    \providecommand\color[2][]{%
      \GenericError{(gnuplot) \space\space\space\@spaces}{%
        Package color not loaded in conjunction with
        terminal option `colourtext'%
      }{See the gnuplot documentation for explanation.%
      }{Either use 'blacktext' in gnuplot or load the package
        color.sty in LaTeX.}%
      \renewcommand\color[2][]{}%
    }%
    \providecommand\includegraphics[2][]{%
      \GenericError{(gnuplot) \space\space\space\@spaces}{%
        Package graphicx or graphics not loaded%
      }{See the gnuplot documentation for explanation.%
      }{The gnuplot epslatex terminal needs graphicx.sty or graphics.sty.}%
      \renewcommand\includegraphics[2][]{}%
    }%
    \providecommand\rotatebox[2]{#2}%
    \@ifundefined{ifGPcolor}{%
      \newif\ifGPcolor
      \GPcolortrue
    }{}%
    \@ifundefined{ifGPblacktext}{%
      \newif\ifGPblacktext
      \GPblacktexttrue
    }{}%
    \let\gplgaddtomacro\g@addto@macro
    \gdef\gplbacktext{}%
    \gdef\gplfronttext{}%
    \makeatother
    \ifGPblacktext
      \def\colorrgb#1{}%
      \def\colorgray#1{}%
    \else
      \ifGPcolor
        \def\colorrgb#1{\color[rgb]{#1}}%
        \def\colorgray#1{\color[gray]{#1}}%
        \expandafter\def\csname LTw\endcsname{\color{white}}%
        \expandafter\def\csname LTb\endcsname{\color{black}}%
        \expandafter\def\csname LTa\endcsname{\color{black}}%
        \expandafter\def\csname LT0\endcsname{\color[rgb]{1,0,0}}%
        \expandafter\def\csname LT1\endcsname{\color[rgb]{0,1,0}}%
        \expandafter\def\csname LT2\endcsname{\color[rgb]{0,0,1}}%
        \expandafter\def\csname LT3\endcsname{\color[rgb]{1,0,1}}%
        \expandafter\def\csname LT4\endcsname{\color[rgb]{0,1,1}}%
        \expandafter\def\csname LT5\endcsname{\color[rgb]{1,1,0}}%
        \expandafter\def\csname LT6\endcsname{\color[rgb]{0,0,0}}%
        \expandafter\def\csname LT7\endcsname{\color[rgb]{1,0.3,0}}%
        \expandafter\def\csname LT8\endcsname{\color[rgb]{0.5,0.5,0.5}}%
      \else
        \def\colorrgb#1{\color{black}}%
        \def\colorgray#1{\color[gray]{#1}}%
        \expandafter\def\csname LTw\endcsname{\color{white}}%
        \expandafter\def\csname LTb\endcsname{\color{black}}%
        \expandafter\def\csname LTa\endcsname{\color{black}}%
        \expandafter\def\csname LT0\endcsname{\color{black}}%
        \expandafter\def\csname LT1\endcsname{\color{black}}%
        \expandafter\def\csname LT2\endcsname{\color{black}}%
        \expandafter\def\csname LT3\endcsname{\color{black}}%
        \expandafter\def\csname LT4\endcsname{\color{black}}%
        \expandafter\def\csname LT5\endcsname{\color{black}}%
        \expandafter\def\csname LT6\endcsname{\color{black}}%
        \expandafter\def\csname LT7\endcsname{\color{black}}%
        \expandafter\def\csname LT8\endcsname{\color{black}}%
      \fi
    \fi
      \setlength{\unitlength}{0.0500bp}%
      \ifx\gptboxheight\undefined%
        \newlength{\gptboxheight}%
        \newlength{\gptboxwidth}%
        \newsavebox{\gptboxtext}%
      \fi%
      \setlength{\fboxrule}{0.5pt}%
      \setlength{\fboxsep}{1pt}%
  \begin{picture}(4800.00,1800.00)%
      \gplgaddtomacro\gplbacktext{%
        \colorrgb{0.00,0.00,0.00}%
        \put(378,594){\makebox(0,0)[r]{\strut{}$\scalebox{0.8}{-1}$}}%
        \colorrgb{0.00,0.00,0.00}%
        \put(378,1170){\makebox(0,0)[r]{\strut{}$\scalebox{0.8}{0}$}}%
        \colorrgb{0.00,0.00,0.00}%
        \put(378,1745){\makebox(0,0)[r]{\strut{}$\scalebox{0.8}{1}$}}%
        \colorrgb{0.00,0.00,0.00}%
        \put(480,408){\makebox(0,0){\strut{}$\scalebox{0.8}{-2}$}}%
        \colorrgb{0.00,0.00,0.00}%
        \put(914,408){\makebox(0,0){\strut{}$\scalebox{0.8}{-1}$}}%
        \colorrgb{0.00,0.00,0.00}%
        \put(1349,408){\makebox(0,0){\strut{}$\scalebox{0.8}{0}$}}%
        \colorrgb{0.00,0.00,0.00}%
        \put(1783,408){\makebox(0,0){\strut{}$\scalebox{0.8}{1}$}}%
        \colorrgb{0.00,0.00,0.00}%
        \put(2217,408){\makebox(0,0){\strut{}$\scalebox{0.8}{2}$}}%
      }%
      \gplgaddtomacro\gplfronttext{%
        \csname LTb\endcsname%
        \put(81,1169){\rotatebox{-270}{\makebox(0,0){\strut{}Current $J/J_c$}}}%
        \csname LTb\endcsname%
        \put(1348,129){\makebox(0,0){\strut{}Phase-difference $\delta\varphi/\pi$}}%
        \colorrgb{0.00,0.00,0.00}%
        \put(378,594){\makebox(0,0)[r]{\strut{}$\scalebox{0.8}{-1}$}}%
        \colorrgb{0.00,0.00,0.00}%
        \put(378,1170){\makebox(0,0)[r]{\strut{}$\scalebox{0.8}{0}$}}%
        \colorrgb{0.00,0.00,0.00}%
        \put(378,1745){\makebox(0,0)[r]{\strut{}$\scalebox{0.8}{1}$}}%
        \colorrgb{0.00,0.00,0.00}%
        \put(480,408){\makebox(0,0){\strut{}$\scalebox{0.8}{-2}$}}%
        \colorrgb{0.00,0.00,0.00}%
        \put(914,408){\makebox(0,0){\strut{}$\scalebox{0.8}{-1}$}}%
        \colorrgb{0.00,0.00,0.00}%
        \put(1349,408){\makebox(0,0){\strut{}$\scalebox{0.8}{0}$}}%
        \colorrgb{0.00,0.00,0.00}%
        \put(1783,408){\makebox(0,0){\strut{}$\scalebox{0.8}{1}$}}%
        \colorrgb{0.00,0.00,0.00}%
        \put(2217,408){\makebox(0,0){\strut{}$\scalebox{0.8}{2}$}}%
        \csname LTb\endcsname%
        \put(502,1630){\makebox(0,0)[l]{\strut{}\textbf{(a)}}}%
      }%
      \gplgaddtomacro\gplbacktext{%
        \colorrgb{0.00,0.00,0.00}%
        \put(2883,594){\makebox(0,0)[r]{\strut{}$\scalebox{0.8}{0}$}}%
        \colorrgb{0.00,0.00,0.00}%
        \put(2883,1170){\makebox(0,0)[r]{\strut{}$\scalebox{0.8}{2}$}}%
        \colorrgb{0.00,0.00,0.00}%
        \put(2883,1745){\makebox(0,0)[r]{\strut{}$\scalebox{0.8}{4}$}}%
        \colorrgb{0.00,0.00,0.00}%
        \put(2985,408){\makebox(0,0){\strut{}$\scalebox{0.8}{-2}$}}%
        \colorrgb{0.00,0.00,0.00}%
        \put(3419,408){\makebox(0,0){\strut{}$\scalebox{0.8}{-1}$}}%
        \colorrgb{0.00,0.00,0.00}%
        \put(3854,408){\makebox(0,0){\strut{}$\scalebox{0.8}{0}$}}%
        \colorrgb{0.00,0.00,0.00}%
        \put(4288,408){\makebox(0,0){\strut{}$\scalebox{0.8}{1}$}}%
        \colorrgb{0.00,0.00,0.00}%
        \put(4722,408){\makebox(0,0){\strut{}$\scalebox{0.8}{2}$}}%
      }%
      \gplgaddtomacro\gplfronttext{%
        \csname LTb\endcsname%
        \put(2596,1169){\rotatebox{-270}{\makebox(0,0){\strut{}Energy $E/E_c$}}}%
        \csname LTb\endcsname%
        \put(3853,129){\makebox(0,0){\strut{}Phase-difference $\delta\varphi/\pi$}}%
        \colorrgb{0.00,0.00,0.00}%
        \put(2883,594){\makebox(0,0)[r]{\strut{}$\scalebox{0.8}{0}$}}%
        \colorrgb{0.00,0.00,0.00}%
        \put(2883,1170){\makebox(0,0)[r]{\strut{}$\scalebox{0.8}{2}$}}%
        \colorrgb{0.00,0.00,0.00}%
        \put(2883,1745){\makebox(0,0)[r]{\strut{}$\scalebox{0.8}{4}$}}%
        \colorrgb{0.00,0.00,0.00}%
        \put(2985,408){\makebox(0,0){\strut{}$\scalebox{0.8}{-2}$}}%
        \colorrgb{0.00,0.00,0.00}%
        \put(3419,408){\makebox(0,0){\strut{}$\scalebox{0.8}{-1}$}}%
        \colorrgb{0.00,0.00,0.00}%
        \put(3854,408){\makebox(0,0){\strut{}$\scalebox{0.8}{0}$}}%
        \colorrgb{0.00,0.00,0.00}%
        \put(4288,408){\makebox(0,0){\strut{}$\scalebox{0.8}{1}$}}%
        \colorrgb{0.00,0.00,0.00}%
        \put(4722,408){\makebox(0,0){\strut{}$\scalebox{0.8}{2}$}}%
        \csname LTb\endcsname%
        \put(2998,1630){\makebox(0,0)[l]{\strut{}\textbf{(b)}}}%
      }%
      \gplbacktext
      \put(0,0){\includegraphics{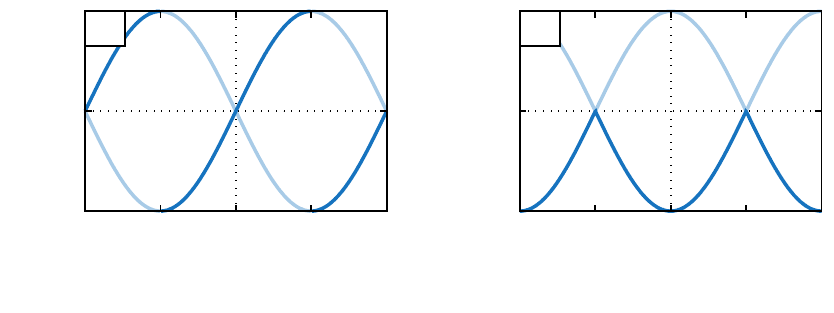}}%
      \gplfronttext
    \end{picture}%
  \endgroup
  \caption
  {%
  Analytical plot of \textbf{(a)} the supercurrent and \textbf{(b)} the free energy as functions of the phase-difference in a $\sin(\delta\varphi/2)$ junction.
  For every value of the phase-difference $\delta\varphi$, there are two solutions: the ground state (dark blue line) and a{n } excited state (transparent line).
  }%
  \label{fig:periodicity}
\end{figure} 

For each phase-difference $\delta\varphi$, one solution {corresponds to the ground state while the other is an excited state,} as illustrated in \cref{fig:periodicity}.
The ground state energy is $2\pi$-periodic{; assuming} the system stays in the ground state as the phase-difference is varied, the full current-phase relation can be written succinctly\cite{DeLuca2009}
\begin{align}
  J = J_c \sin(\delta\varphi/2) \,\sgn[\cos(\delta\varphi/2)] \,.
\end{align}
Throughout this manuscript, we will for brevity refer to the current-phase relation as having a $\sin(\delta\varphi/2)$ \emph{shape}, even though it is {$2\pi$-periodic due to discontinuities at $\delta\varphi=\pm\pi$}.

The critical current $J_c$ obtained for the S/X/S/X/S junction is the same as for each of the S/X/S subjunctions, and we would in fact obtain the same result for a general $\sin(\delta\varphi/n)$ junction.
However, one subtlety is that the \emph{ground state current} is capped at $J_c \sin(\pi/n)$ due to the discontinuities at $\delta\varphi = \pm\pi$; higher currents correspond to the excited states.
In practice, this should not be a problem, as even the ground state current is reasonably robust with respect to the number of barriers: the maximal ground state current is still as large as $J_c/2$ for $n=6$, and does not drop below $J_c/10$ before $n=32$.


{
With this in mind, let us consider S/F/S/F/S double-barrier junctions, where the central F/S/F acts as a superconducting spin-valve (see \cref{fig:model}).
{Switching the orientation of only one F in such a structure can \eg be achieved using materials with intrinsically different coercivities\cite{Gingrich2016}, or using different thicknesses for the ferromagnets, where the latter affects both coercivities and anisotropies.}
Depending on whether the two F layers have a parallel (P) or antiparallel (AP) orientation, their effective magnetic exchange fields induced in the central S will either add or cancel. 
The stability of the superconducting condensate depends on the net exchange field felt by the Cooper pairs there, which means that one can toggle superconductivity on and off in the central region \cite{degennes_pl_66, hauser_prl_69}, effectively switching between an S/X/S and S/X/S/X/S junction.
Since the S/X/S junction {always has a $\sin(\delta\varphi)$-shaped current-phase relation,} while an S/X/S/X/S junction {also supports $\sin(\delta\varphi/2)$-shaped solutions}, the result is that one should be able to magnetically switch between these current-phase relations.
In \cref{sec:numerical}, we show the results of extensive selfconsistent numerical simulations which confirm this prediction.
}

\subsection{{Triple-barrier junction}}\label{sec:sinphi3}\noindent
Let us now move on to a slightly more complicated case, namely S/X/S/X/S/X/S junctions.
In other words, we now have three concatenated S/X/S junctions, which we can describe using three currents $J_{12},\ldots,J_{34}$ and four phases $\varphi_1,\ldots,\varphi_4$:
\begin{align}
  J_{12} &= J_c \sin(\varphi_2 - \varphi_1), \\
  J_{23} &= J_c \sin(\varphi_3 - \varphi_2), \\
  J_{34} &= J_c \sin(\varphi_4 - \varphi_3).
\end{align}
To simplify the derivations, we parametrize the superconductor phases $\varphi_1,\ldots,\varphi_4$ in terms of their averages and differences:
\begin{align}
  \varphi_1 &\equiv a - d/2,  &
  \varphi_2 &\equiv b - c/2,  \\
  \varphi_4 &\equiv a + d/2,  &
  \varphi_3 &\equiv b + c/2.
\end{align}
In other words, the outer superconductors are described by their average~$a$ and difference~$d$; while the inner ones are described by the corresponding parameters $b$ and $c$.
Since the overall phase of the system has no physical significance, we can set $a=0$ without loss of generality.
Thus, we obtain:
\begin{align}
  J_{12} &= J_c \sin(+b - c/2 + d/2), \\
  J_{23} &= J_c \sin(c), \\
  J_{34} &= J_c \sin(-b - c/2 + d/2).
\end{align}

\begin{figure}[b] 
  \centering
  \begingroup
    \makeatletter
    \providecommand\color[2][]{%
      \GenericError{(gnuplot) \space\space\space\@spaces}{%
        Package color not loaded in conjunction with
        terminal option `colourtext'%
      }{See the gnuplot documentation for explanation.%
      }{Either use 'blacktext' in gnuplot or load the package
        color.sty in LaTeX.}%
      \renewcommand\color[2][]{}%
    }%
    \providecommand\includegraphics[2][]{%
      \GenericError{(gnuplot) \space\space\space\@spaces}{%
        Package graphicx or graphics not loaded%
      }{See the gnuplot documentation for explanation.%
      }{The gnuplot epslatex terminal needs graphicx.sty or graphics.sty.}%
      \renewcommand\includegraphics[2][]{}%
    }%
    \providecommand\rotatebox[2]{#2}%
    \@ifundefined{ifGPcolor}{%
      \newif\ifGPcolor
      \GPcolortrue
    }{}%
    \@ifundefined{ifGPblacktext}{%
      \newif\ifGPblacktext
      \GPblacktexttrue
    }{}%
    \let\gplgaddtomacro\g@addto@macro
    \gdef\gplbacktext{}%
    \gdef\gplfronttext{}%
    \makeatother
    \ifGPblacktext
      \def\colorrgb#1{}%
      \def\colorgray#1{}%
    \else
      \ifGPcolor
        \def\colorrgb#1{\color[rgb]{#1}}%
        \def\colorgray#1{\color[gray]{#1}}%
        \expandafter\def\csname LTw\endcsname{\color{white}}%
        \expandafter\def\csname LTb\endcsname{\color{black}}%
        \expandafter\def\csname LTa\endcsname{\color{black}}%
        \expandafter\def\csname LT0\endcsname{\color[rgb]{1,0,0}}%
        \expandafter\def\csname LT1\endcsname{\color[rgb]{0,1,0}}%
        \expandafter\def\csname LT2\endcsname{\color[rgb]{0,0,1}}%
        \expandafter\def\csname LT3\endcsname{\color[rgb]{1,0,1}}%
        \expandafter\def\csname LT4\endcsname{\color[rgb]{0,1,1}}%
        \expandafter\def\csname LT5\endcsname{\color[rgb]{1,1,0}}%
        \expandafter\def\csname LT6\endcsname{\color[rgb]{0,0,0}}%
        \expandafter\def\csname LT7\endcsname{\color[rgb]{1,0.3,0}}%
        \expandafter\def\csname LT8\endcsname{\color[rgb]{0.5,0.5,0.5}}%
      \else
        \def\colorrgb#1{\color{black}}%
        \def\colorgray#1{\color[gray]{#1}}%
        \expandafter\def\csname LTw\endcsname{\color{white}}%
        \expandafter\def\csname LTb\endcsname{\color{black}}%
        \expandafter\def\csname LTa\endcsname{\color{black}}%
        \expandafter\def\csname LT0\endcsname{\color{black}}%
        \expandafter\def\csname LT1\endcsname{\color{black}}%
        \expandafter\def\csname LT2\endcsname{\color{black}}%
        \expandafter\def\csname LT3\endcsname{\color{black}}%
        \expandafter\def\csname LT4\endcsname{\color{black}}%
        \expandafter\def\csname LT5\endcsname{\color{black}}%
        \expandafter\def\csname LT6\endcsname{\color{black}}%
        \expandafter\def\csname LT7\endcsname{\color{black}}%
        \expandafter\def\csname LT8\endcsname{\color{black}}%
      \fi
    \fi
      \setlength{\unitlength}{0.0500bp}%
      \ifx\gptboxheight\undefined%
        \newlength{\gptboxheight}%
        \newlength{\gptboxwidth}%
        \newsavebox{\gptboxtext}%
      \fi%
      \setlength{\fboxrule}{0.5pt}%
      \setlength{\fboxsep}{1pt}%
  \begin{picture}(4800.00,1800.00)%
      \gplgaddtomacro\gplbacktext{%
        \colorrgb{0.00,0.00,0.00}%
        \put(378,594){\makebox(0,0)[r]{\strut{}$\scalebox{0.8}{-1}$}}%
        \colorrgb{0.00,0.00,0.00}%
        \put(378,1170){\makebox(0,0)[r]{\strut{}$\scalebox{0.8}{0}$}}%
        \colorrgb{0.00,0.00,0.00}%
        \put(378,1745){\makebox(0,0)[r]{\strut{}$\scalebox{0.8}{1}$}}%
        \colorrgb{0.00,0.00,0.00}%
        \put(480,408){\makebox(0,0){\strut{}$\scalebox{0.8}{-3}$}}%
        \colorrgb{0.00,0.00,0.00}%
        \put(769,408){\makebox(0,0){\strut{}$\scalebox{0.8}{-2}$}}%
        \colorrgb{0.00,0.00,0.00}%
        \put(1059,408){\makebox(0,0){\strut{}$\scalebox{0.8}{-1}$}}%
        \colorrgb{0.00,0.00,0.00}%
        \put(1349,408){\makebox(0,0){\strut{}$\scalebox{0.8}{0}$}}%
        \colorrgb{0.00,0.00,0.00}%
        \put(1638,408){\makebox(0,0){\strut{}$\scalebox{0.8}{1}$}}%
        \colorrgb{0.00,0.00,0.00}%
        \put(1928,408){\makebox(0,0){\strut{}$\scalebox{0.8}{2}$}}%
        \colorrgb{0.00,0.00,0.00}%
        \put(2217,408){\makebox(0,0){\strut{}$\scalebox{0.8}{3}$}}%
      }%
      \gplgaddtomacro\gplfronttext{%
        \csname LTb\endcsname%
        \put(81,1169){\rotatebox{-270}{\makebox(0,0){\strut{}Current $J/J_c$}}}%
        \csname LTb\endcsname%
        \put(1348,129){\makebox(0,0){\strut{}Phase-difference $\delta\varphi/\pi$}}%
        \colorrgb{0.00,0.00,0.00}%
        \put(378,594){\makebox(0,0)[r]{\strut{}$\scalebox{0.8}{-1}$}}%
        \colorrgb{0.00,0.00,0.00}%
        \put(378,1170){\makebox(0,0)[r]{\strut{}$\scalebox{0.8}{0}$}}%
        \colorrgb{0.00,0.00,0.00}%
        \put(378,1745){\makebox(0,0)[r]{\strut{}$\scalebox{0.8}{1}$}}%
        \colorrgb{0.00,0.00,0.00}%
        \put(480,408){\makebox(0,0){\strut{}$\scalebox{0.8}{-3}$}}%
        \colorrgb{0.00,0.00,0.00}%
        \put(769,408){\makebox(0,0){\strut{}$\scalebox{0.8}{-2}$}}%
        \colorrgb{0.00,0.00,0.00}%
        \put(1059,408){\makebox(0,0){\strut{}$\scalebox{0.8}{-1}$}}%
        \colorrgb{0.00,0.00,0.00}%
        \put(1349,408){\makebox(0,0){\strut{}$\scalebox{0.8}{0}$}}%
        \colorrgb{0.00,0.00,0.00}%
        \put(1638,408){\makebox(0,0){\strut{}$\scalebox{0.8}{1}$}}%
        \colorrgb{0.00,0.00,0.00}%
        \put(1928,408){\makebox(0,0){\strut{}$\scalebox{0.8}{2}$}}%
        \colorrgb{0.00,0.00,0.00}%
        \put(2217,408){\makebox(0,0){\strut{}$\scalebox{0.8}{3}$}}%
        \csname LTb\endcsname%
        \put(499,1630){\makebox(0,0)[l]{\strut{}\textbf{(a)}}}%
      }%
      \gplgaddtomacro\gplbacktext{%
        \colorrgb{0.00,0.00,0.00}%
        \put(2883,594){\makebox(0,0)[r]{\strut{}$\scalebox{0.8}{0}$}}%
        \colorrgb{0.00,0.00,0.00}%
        \put(2883,1170){\makebox(0,0)[r]{\strut{}$\scalebox{0.8}{3}$}}%
        \colorrgb{0.00,0.00,0.00}%
        \put(2883,1745){\makebox(0,0)[r]{\strut{}$\scalebox{0.8}{6}$}}%
        \colorrgb{0.00,0.00,0.00}%
        \put(2985,408){\makebox(0,0){\strut{}$\scalebox{0.8}{-3}$}}%
        \colorrgb{0.00,0.00,0.00}%
        \put(3274,408){\makebox(0,0){\strut{}$\scalebox{0.8}{-2}$}}%
        \colorrgb{0.00,0.00,0.00}%
        \put(3564,408){\makebox(0,0){\strut{}$\scalebox{0.8}{-1}$}}%
        \colorrgb{0.00,0.00,0.00}%
        \put(3854,408){\makebox(0,0){\strut{}$\scalebox{0.8}{0}$}}%
        \colorrgb{0.00,0.00,0.00}%
        \put(4143,408){\makebox(0,0){\strut{}$\scalebox{0.8}{1}$}}%
        \colorrgb{0.00,0.00,0.00}%
        \put(4433,408){\makebox(0,0){\strut{}$\scalebox{0.8}{2}$}}%
        \colorrgb{0.00,0.00,0.00}%
        \put(4722,408){\makebox(0,0){\strut{}$\scalebox{0.8}{3}$}}%
      }%
      \gplgaddtomacro\gplfronttext{%
        \csname LTb\endcsname%
        \put(2596,1169){\rotatebox{-270}{\makebox(0,0){\strut{}Energy $E/E_c$}}}%
        \csname LTb\endcsname%
        \put(3853,129){\makebox(0,0){\strut{}Phase-difference $\delta\varphi/\pi$}}%
        \colorrgb{0.00,0.00,0.00}%
        \put(2883,594){\makebox(0,0)[r]{\strut{}$\scalebox{0.8}{0}$}}%
        \colorrgb{0.00,0.00,0.00}%
        \put(2883,1170){\makebox(0,0)[r]{\strut{}$\scalebox{0.8}{3}$}}%
        \colorrgb{0.00,0.00,0.00}%
        \put(2883,1745){\makebox(0,0)[r]{\strut{}$\scalebox{0.8}{6}$}}%
        \colorrgb{0.00,0.00,0.00}%
        \put(2985,408){\makebox(0,0){\strut{}$\scalebox{0.8}{-3}$}}%
        \colorrgb{0.00,0.00,0.00}%
        \put(3274,408){\makebox(0,0){\strut{}$\scalebox{0.8}{-2}$}}%
        \colorrgb{0.00,0.00,0.00}%
        \put(3564,408){\makebox(0,0){\strut{}$\scalebox{0.8}{-1}$}}%
        \colorrgb{0.00,0.00,0.00}%
        \put(3854,408){\makebox(0,0){\strut{}$\scalebox{0.8}{0}$}}%
        \colorrgb{0.00,0.00,0.00}%
        \put(4143,408){\makebox(0,0){\strut{}$\scalebox{0.8}{1}$}}%
        \colorrgb{0.00,0.00,0.00}%
        \put(4433,408){\makebox(0,0){\strut{}$\scalebox{0.8}{2}$}}%
        \colorrgb{0.00,0.00,0.00}%
        \put(4722,408){\makebox(0,0){\strut{}$\scalebox{0.8}{3}$}}%
        \csname LTb\endcsname%
        \put(2999,1630){\makebox(0,0)[l]{\strut{}\textbf{(b)}}}%
      }%
      \gplbacktext
      \put(0,0){\includegraphics{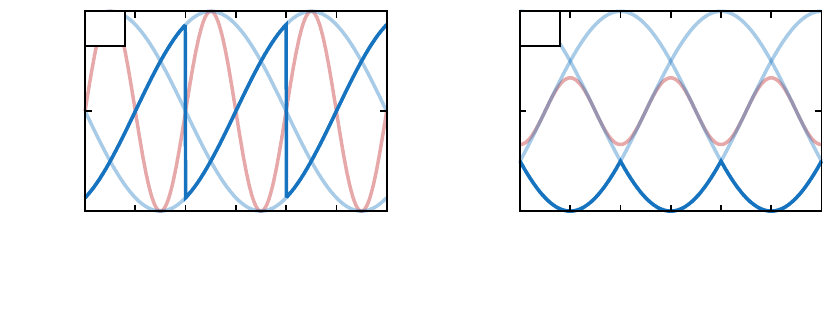}}%
      \gplfronttext
    \end{picture}%
  \endgroup
  \caption
  {%
  Analytical plot of \textbf{(a)} the supercurrent and \textbf{(b)} the free energy as functions of the phase-difference in a $\sin(\delta\varphi/3)$ junction.
  There are multiple solutions: the $\sin(\delta\varphi/3)$-shaped ground state (dark blue line), the $\sin(\delta\varphi/3)$-shaped excited states (transparent blue lines), and the $\sin(\delta\varphi)$-shaped excited state (transparent red line).
  Note that the ground state is $2\pi$-periodic due to discontinuities at $\delta\varphi = \pm\pi, \pm 3\pi$.
  }%
  \label{fig:periodicity3}
\end{figure} 

To determine the current-phase relation of the junction, we need to solve the current conservation equation $J_{12} = J_{23} = J_{34}$ for a fixed external phase-difference $\delta\varphi \equiv d \equiv \varphi_4 - \varphi_1$.
Let us first consider the part $J_{12} = J_{34}$.
Invoking the trigonometric identity $\sin(u \pm v) = \sin(u)\cos(v) \pm \cos(u)\sin(v)$ on both sides of the equation, this can be rewritten as $\sin(b) \cos(d/2 - c/2) = 0$.
This has three distinct solutions: $b=0$; $b = \pi$; and $c = d + \pi$.
We will first address the solution branches ${b = 0}$ and ${b = \pi}$.
If we substitute these solutions into $J_{12} = J_{23}$, we obtain:
\begin{equation}
  \sin(c) = \pm\sin(d/2 - c/2) .
\end{equation}
This equation admits four distinct solutions: ${b = 0}$, ${c = d/3}$; ${b = \pi}$, ${c = (d-2\pi)/3}$; ${b = \pi}$, ${c = (d+2\pi)/3}$; ${b = \pi}$, ${c = -d}$.
Let us first consider the first three solutions.
Substituting these back into the currents and energies of each subjunction, restoring the notation $\delta\varphi \equiv d$, and defining $k\in\{-1,0,+1\}$, we find the current-phase and energy-phase relations:
\begin{equation}
\begin{aligned}
  J &= J_c \sin\big[(\delta\varphi + 2\pi k)/3\big], \\
  E &= E_c\!\left\{ 3 - 3\cos\big[(\delta\varphi + 2 \pi k)/3\big]\right\}.
\end{aligned}
\end{equation}
In other words, we do indeed find three solutions where the current-phase relation has the shape $\sin(\delta\varphi/3)$, as expected.

During the derivation above, we pointed out two alternative ways to satisfy the equations.
Substituting these into the current conservation equation, we identify three additional solutions: ${b = \pi}$, ${c = -d}$; ${b = d + \pi/2}$, ${c = d+\pi}$; ${b = d - \pi/2}$, ${c = d+\pi}$.
The currents and energies of these solutions are identical:
\begin{align}
  J &= -J_c \sin(\delta\varphi), &
  E &= E_c [3 + \cos(\delta\varphi)].
\end{align}
In other words, the alternative solutions correspond to $\sin(\delta\varphi)${-shaped current-phase relations in the $\pi$-phase}.
However, as shown in \cref{fig:periodicity3}, these solutions are always excited states, and cannot be accessed in a system that relaxes quickly to the ground state.
The reason is that these alternative solution branches have two subjunctions with the energy-phase relation $E = E_c[1+\cos(\delta\varphi)]$ and one with $E=E_c[1-\cos(\delta\varphi)]$, meaning that minimizing the energy of one subjunction will simultaneously maximize the energy of another.
This is different from the $\sin(\delta\varphi/3)${-shaped} solutions above, where the energy minima of each subjunction coincide.

\subsection{{Generalization}}\label{sec:sinphin}\noindent
Let us now consider a general $n$-barrier junction, \ie a system of $n+1$ identical superconductors separated by $n$ identical low-transparency barriers.
As in the previous subsections, we can associate one phase $\varphi_0,\ldots,\varphi_n$ with each superconductor, and write the currents and energies of the $n$ subjunctions as
\begin{align}
  J_{m,m+1} &= J_c \sin(\varphi_{m+1}    - \varphi_m), \\
  E_{m,m+1} &= E_c [1-\cos(\varphi_{m+1} - \varphi_m)]. 
\end{align}
We fix the phase of the first superconductor $\varphi_0 = 0$, and the last one $\varphi_n = \delta\varphi$, so that $\delta\varphi$ is the phase-difference between the outer superconductors.
It is then straight-forward to verify that the following phase-distribution satisfies these requirements (modulo $2\pi$), while also ensuring current conservation:
\begin{align}
  \varphi_m &= (m/n)(\delta\varphi+2\pi k),
\end{align}
where $m \in \{0,1,\ldots,n\}$ identifies each superconductor, and $k \in \{0,1,\ldots,n-1\}$ identifies each solution branch.
Substituted into the current-phase and energy-phase relations of each subjunction, we get the following equations for the entire junction:
\begin{align}
  J &= J_c  \sin\big[(\delta\varphi+2\pi k)/n\big], \\
  E &= E_c \left\{n - n\cos\big[(\delta\varphi+2\pi k)/n\big]\right\}.
\end{align}
Each of these solution branches has to contribute to the ground state current-phase relation, as each of them provides a global energy minimum $E = 0$ for some phase-difference~$\delta\varphi$.
If the system stays in the lowest-energy configuration as the phase-difference is varied, we again recover $2\pi$-periodicity.

The current conservation equations may have more solutions than the ones presented above, as we discovered in \cref{sec:sinphi3} for the special case of a triple-barrier junction.
However, in the triple-barrier case, we also found that these solutions corresponded to excited states, because the energy minima and maxima of each subjunction occured at different phase-differences.
One might expect such a pattern to persist for $n>3$, in which case $n$-barrier junctions would always have $2\pi$-periodic $\sin(\delta\varphi/n)${-shaped} current-phase relations {with discontinuities at $\delta\varphi=\pm\pi$ as their} ground state.


\section{Numerical results}\label{sec:numerical}\noindent
To investigate the idea{s} outlined {in the previous sections}, we numerically investigated the simplest possible structure of this kind, namely an S/F/S/F/S structure where the ferromagnetic layers are atomically thin insulators.
The superconductors at the ends of the junction are assumed to be much larger than the superconducting coherence length $\xi$, so that we can treat them as bulk superconductors with order parameters $\Delta_0 e^{\pm \i\delta\varphi/2}$, where $\Delta_0$ and $\delta\varphi$ are real numbers.
The central superconductor is assumed to have the length $L = \xi/2$.
To describe the superconducting correlations there properly, we had to simultaneously solve the Usadel diffusion equation~\cite{usadel_prl_70},
\begin{equation}
  \i \hbar D \partial_z (\hat{g} \partial_z \hat{g}) = [(\epsilon+\i\delta)\hat{\tau}_3 + \hat{\Delta}, \hat{g}],
\end{equation}
and the gap equation, 
\begin{equation}
  \Delta(z) = \frac{1}{2}N_0\lambda\! \int^{+\Theta}_{-\Theta} \!\!\d\epsilon\, f_s(z,\epsilon) \tanh(\epsilon/2T),
\end{equation}
yielding selfconsistent results. Above, $\Theta=\Delta_0\cosh(1/N_0\lambda)$ is the Debye cutoff, $\epsilon$ the quasiparticle energy, $\delta$ the inelastic scattering rate, $N_0$ the normal-state density of states at the Fermi level, $\lambda$ the BCS coupling constant, $D = \Delta_0\xi^{2\!}/\hbar$ the diffusion coefficient,  $T$ the temperature of the system, and $\hbar$ Planck's reduced constant. The $4\times4$ matrix $\hat{g}(z,\epsilon)$ contains the spin-resolved normal and anomalous retarded Green functions in the system, ${\hat{\Delta} = \text{antidiag}(+\Delta,-\Delta,+\Delta^{\!*},-\Delta^{\!*})}$ contains the superconducting order parameter $\Delta(z)$, and ${\hat{\tau}_3 = \text{diag}(+1,+1,-1,-1)}$.
Only the spin-singlet part $f_s$ of the anomalous Green function enters the gap equation above.
For more details about the numerical solution of these equations, see Refs.~\onlinecite{Ouassou2016a}~and~\onlinecite{Ouassou2016b}.

The ferromagnetic insulators are approximated as spin-active interfaces between the superconducting layers, which we describe using the recently derived spin-dependent boundary conditions for strongly polarized interfaces \cite{Eschrig2015}. 
{The boundary conditions can be written in terms of the \emph{tunneling conductance}~$G_T$, which depends on the number of transmission channels and their transparencies; a \emph{spin-mixing conductance}~$G_\varphi$, which describes the difference in phase-shift between spin-up and spin-down electrons reflected off the interface; and a spin-polarization~$P$, which describes the different interface transparencies perceived by spin-up and spin-down electrons.
See Refs.~\onlinecite{Eschrig2015}~and~\onlinecite{Ouassou2016b} for more details about these parameters.}
We assumed a strong interface polarization $P = 0.90$, a small tunneling conductance $G_T = 0.02G_0$, and a spin-mixing conductance $G_\varphi = 0.3G_0$, where $G_0$ is the normal-state conductance of the central superconductor.
Such a parameter choice is likely suitable for strongly polarized ferromagnetic insulators such as GdN or EuS, where polarizations of 90\% and upward have been experimentally reported \cite{li_prl_13, pal_prb_15}.
Finally, we calculated the supercurrent throughout the junction via the equation
\begin{equation}
  J(z) = 2J_0\! \int_{-\Theta}^{+\Theta} \!\!\d\epsilon\, \mathrm{Re}\,\mathrm{Tr} \big[ \hat{\tau}_{3\,} \hat{g}(z,\epsilon)\, \partial_{z\,} \hat{g}(z,\epsilon) \big] \tanh(\epsilon/2T),
\end{equation}
where ${J_0^{\vphantom{2}} = eN_0^{\vphantom{2}} \Delta_0^2 \xi^2 \!A/ 4\hbar L}$, $e$ is the electron charge, and $A$ is the cross-sectional area of the central superconductor.

\subsection{Ideal case: symmetric junction}\label{sec:num-sinphi2-symm}
\begin{figure}[t!] 
  \centering
    \begingroup
      \makeatletter
      \providecommand\color[2][]{%
        \GenericError{(gnuplot) \space\space\space\@spaces}{%
          Package color not loaded in conjunction with
          terminal option `colourtext'%
        }{See the gnuplot documentation for explanation.%
        }{Either use 'blacktext' in gnuplot or load the package
          color.sty in LaTeX.}%
        \renewcommand\color[2][]{}%
      }%
      \providecommand\includegraphics[2][]{%
        \GenericError{(gnuplot) \space\space\space\@spaces}{%
          Package graphicx or graphics not loaded%
        }{See the gnuplot documentation for explanation.%
        }{The gnuplot epslatex terminal needs graphicx.sty or graphics.sty.}%
        \renewcommand\includegraphics[2][]{}%
      }%
      \providecommand\rotatebox[2]{#2}%
      \@ifundefined{ifGPcolor}{%
        \newif\ifGPcolor
        \GPcolortrue
      }{}%
      \@ifundefined{ifGPblacktext}{%
        \newif\ifGPblacktext
        \GPblacktexttrue
      }{}%
      \let\gplgaddtomacro\g@addto@macro
      \gdef\gplbacktext{}%
      \gdef\gplfronttext{}%
      \makeatother
      \ifGPblacktext
        \def\colorrgb#1{}%
        \def\colorgray#1{}%
      \else
        \ifGPcolor
          \def\colorrgb#1{\color[rgb]{#1}}%
          \def\colorgray#1{\color[gray]{#1}}%
          \expandafter\def\csname LTw\endcsname{\color{white}}%
          \expandafter\def\csname LTb\endcsname{\color{black}}%
          \expandafter\def\csname LTa\endcsname{\color{black}}%
          \expandafter\def\csname LT0\endcsname{\color[rgb]{1,0,0}}%
          \expandafter\def\csname LT1\endcsname{\color[rgb]{0,1,0}}%
          \expandafter\def\csname LT2\endcsname{\color[rgb]{0,0,1}}%
          \expandafter\def\csname LT3\endcsname{\color[rgb]{1,0,1}}%
          \expandafter\def\csname LT4\endcsname{\color[rgb]{0,1,1}}%
          \expandafter\def\csname LT5\endcsname{\color[rgb]{1,1,0}}%
          \expandafter\def\csname LT6\endcsname{\color[rgb]{0,0,0}}%
          \expandafter\def\csname LT7\endcsname{\color[rgb]{1,0.3,0}}%
          \expandafter\def\csname LT8\endcsname{\color[rgb]{0.5,0.5,0.5}}%
        \else
          \def\colorrgb#1{\color{black}}%
          \def\colorgray#1{\color[gray]{#1}}%
          \expandafter\def\csname LTw\endcsname{\color{white}}%
          \expandafter\def\csname LTb\endcsname{\color{black}}%
          \expandafter\def\csname LTa\endcsname{\color{black}}%
          \expandafter\def\csname LT0\endcsname{\color{black}}%
          \expandafter\def\csname LT1\endcsname{\color{black}}%
          \expandafter\def\csname LT2\endcsname{\color{black}}%
          \expandafter\def\csname LT3\endcsname{\color{black}}%
          \expandafter\def\csname LT4\endcsname{\color{black}}%
          \expandafter\def\csname LT5\endcsname{\color{black}}%
          \expandafter\def\csname LT6\endcsname{\color{black}}%
          \expandafter\def\csname LT7\endcsname{\color{black}}%
          \expandafter\def\csname LT8\endcsname{\color{black}}%
        \fi
      \fi
        \setlength{\unitlength}{0.0500bp}%
        \ifx\gptboxheight\undefined%
          \newlength{\gptboxheight}%
          \newlength{\gptboxwidth}%
          \newsavebox{\gptboxtext}%
        \fi%
        \setlength{\fboxrule}{0.5pt}%
        \setlength{\fboxsep}{1pt}%
    \begin{picture}(4520.00,2820.00)%
        \gplgaddtomacro\gplbacktext{%
          \colorrgb{0.00,0.00,0.00}%
          \put(543,631){\makebox(0,0)[r]{\strut{}-1}}%
          \colorrgb{0.00,0.00,0.00}%
          \put(543,1543){\makebox(0,0)[r]{\strut{}0}}%
          \colorrgb{0.00,0.00,0.00}%
          \put(543,2455){\makebox(0,0)[r]{\strut{}1}}%
          \colorrgb{0.00,0.00,0.00}%
          \put(645,409){\makebox(0,0){\strut{}-1}}%
          \colorrgb{0.00,0.00,0.00}%
          \put(2429,409){\makebox(0,0){\strut{}0}}%
          \colorrgb{0.00,0.00,0.00}%
          \put(4213,409){\makebox(0,0){\strut{}1}}%
        }%
        \gplgaddtomacro\gplfronttext{%
          \csname LTb\endcsname%
          \put(144,1543){\rotatebox{-270}{\makebox(0,0){\strut{}Current $J(z)/J_c$}}}%
          \csname LTb\endcsname%
          \put(2429,130){\makebox(0,0){\strut{}Phase-difference $\delta\varphi/\pi$}}%
          \csname LTb\endcsname%
          \put(2083,2684){\makebox(0,0)[r]{\strut{}               AP  [$J_c = J_0/4$]}}%
          \csname LTb\endcsname%
          \put(3721,2684){\makebox(0,0)[r]{\strut{}  P  [$J_c = J_0/200$]}}%
          \colorrgb{0.00,0.00,0.00}%
          \put(543,631){\makebox(0,0)[r]{\strut{}-1}}%
          \colorrgb{0.00,0.00,0.00}%
          \put(543,1543){\makebox(0,0)[r]{\strut{}0}}%
          \colorrgb{0.00,0.00,0.00}%
          \put(543,2455){\makebox(0,0)[r]{\strut{}1}}%
          \colorrgb{0.00,0.00,0.00}%
          \put(645,409){\makebox(0,0){\strut{}-1}}%
          \colorrgb{0.00,0.00,0.00}%
          \put(2429,409){\makebox(0,0){\strut{}0}}%
          \colorrgb{0.00,0.00,0.00}%
          \put(4213,409){\makebox(0,0){\strut{}1}}%
        }%
        \gplbacktext
        \put(0,0){\includegraphics{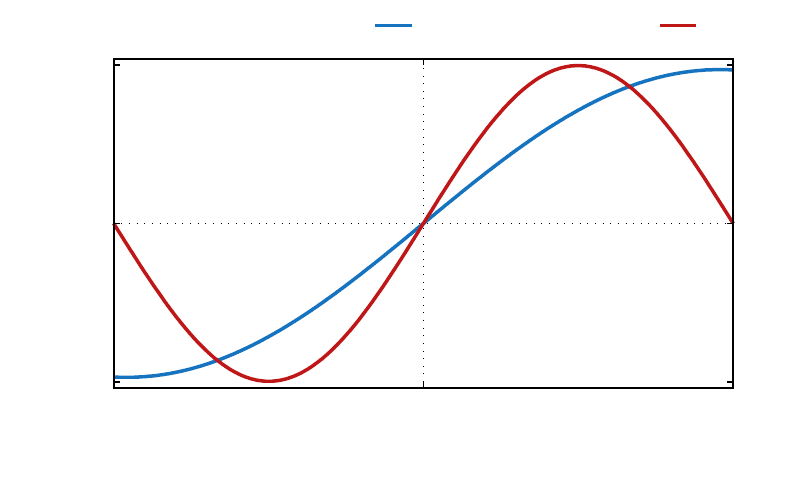}}%
        \gplfronttext
      \end{picture}%
    \endgroup
  \caption
  {%
    Numerical results for the current-phase relation of an S/F/S/F/S junction.
    When we flip the magnetization direction of one of the magnets, we clearly switch between a $J = J_c \sin(\delta\varphi/2)$ and $J = J_c \sin(\delta\varphi)$ current-phase relation.
    Note that the the magnitude of the critical current $J_c$ is rougly 50 times larger in the AP orientation compared to the P orientation, as indicated in the figure legend.
  }%
  \label{fig:cpr}
\end{figure}
\begin{figure}[t!] 
  \centering
      \begingroup
        \makeatletter
        \providecommand\color[2][]{%
          \GenericError{(gnuplot) \space\space\space\@spaces}{%
            Package color not loaded in conjunction with
            terminal option `colourtext'%
          }{See the gnuplot documentation for explanation.%
          }{Either use 'blacktext' in gnuplot or load the package
            color.sty in LaTeX.}%
          \renewcommand\color[2][]{}%
        }%
        \providecommand\includegraphics[2][]{%
          \GenericError{(gnuplot) \space\space\space\@spaces}{%
            Package graphicx or graphics not loaded%
          }{See the gnuplot documentation for explanation.%
          }{The gnuplot epslatex terminal needs graphicx.sty or graphics.sty.}%
          \renewcommand\includegraphics[2][]{}%
        }%
        \providecommand\rotatebox[2]{#2}%
        \@ifundefined{ifGPcolor}{%
          \newif\ifGPcolor
          \GPcolortrue
        }{}%
        \@ifundefined{ifGPblacktext}{%
          \newif\ifGPblacktext
          \GPblacktexttrue
        }{}%
        \let\gplgaddtomacro\g@addto@macro
        \gdef\gplbacktext{}%
        \gdef\gplfronttext{}%
        \makeatother
        \ifGPblacktext
          \def\colorrgb#1{}%
          \def\colorgray#1{}%
        \else
          \ifGPcolor
            \def\colorrgb#1{\color[rgb]{#1}}%
            \def\colorgray#1{\color[gray]{#1}}%
            \expandafter\def\csname LTw\endcsname{\color{white}}%
            \expandafter\def\csname LTb\endcsname{\color{black}}%
            \expandafter\def\csname LTa\endcsname{\color{black}}%
            \expandafter\def\csname LT0\endcsname{\color[rgb]{1,0,0}}%
            \expandafter\def\csname LT1\endcsname{\color[rgb]{0,1,0}}%
            \expandafter\def\csname LT2\endcsname{\color[rgb]{0,0,1}}%
            \expandafter\def\csname LT3\endcsname{\color[rgb]{1,0,1}}%
            \expandafter\def\csname LT4\endcsname{\color[rgb]{0,1,1}}%
            \expandafter\def\csname LT5\endcsname{\color[rgb]{1,1,0}}%
            \expandafter\def\csname LT6\endcsname{\color[rgb]{0,0,0}}%
            \expandafter\def\csname LT7\endcsname{\color[rgb]{1,0.3,0}}%
            \expandafter\def\csname LT8\endcsname{\color[rgb]{0.5,0.5,0.5}}%
          \else
            \def\colorrgb#1{\color{black}}%
            \def\colorgray#1{\color[gray]{#1}}%
            \expandafter\def\csname LTw\endcsname{\color{white}}%
            \expandafter\def\csname LTb\endcsname{\color{black}}%
            \expandafter\def\csname LTa\endcsname{\color{black}}%
            \expandafter\def\csname LT0\endcsname{\color{black}}%
            \expandafter\def\csname LT1\endcsname{\color{black}}%
            \expandafter\def\csname LT2\endcsname{\color{black}}%
            \expandafter\def\csname LT3\endcsname{\color{black}}%
            \expandafter\def\csname LT4\endcsname{\color{black}}%
            \expandafter\def\csname LT5\endcsname{\color{black}}%
            \expandafter\def\csname LT6\endcsname{\color{black}}%
            \expandafter\def\csname LT7\endcsname{\color{black}}%
            \expandafter\def\csname LT8\endcsname{\color{black}}%
          \fi
        \fi
          \setlength{\unitlength}{0.0500bp}%
          \ifx\gptboxheight\undefined%
            \newlength{\gptboxheight}%
            \newlength{\gptboxwidth}%
            \newsavebox{\gptboxtext}%
          \fi%
          \setlength{\fboxrule}{0.5pt}%
          \setlength{\fboxsep}{1pt}%
      \begin{picture}(4800.00,3960.00)%
          \gplgaddtomacro\gplbacktext{%
            \colorrgb{0.00,0.00,0.00}%
            \put(666,2237){\makebox(0,0)[r]{\strut{}$\scalebox{0.8}{0.000}$}}%
            \colorrgb{0.00,0.00,0.00}%
            \put(666,2940){\makebox(0,0)[r]{\strut{}$\scalebox{0.8}{0.020}$}}%
            \colorrgb{0.00,0.00,0.00}%
            \put(666,3642){\makebox(0,0)[r]{\strut{}$\scalebox{0.8}{0.040}$}}%
            \colorrgb{0.00,0.00,0.00}%
            \put(768,2051){\makebox(0,0){\strut{}}}%
            \colorrgb{0.00,0.00,0.00}%
            \put(1617,2051){\makebox(0,0){\strut{}}}%
            \colorrgb{0.00,0.00,0.00}%
            \put(2466,2051){\makebox(0,0){\strut{}}}%
            \csname LTb\endcsname%
            \put(819,3502){\makebox(0,0)[l]{\strut{}\textbf{(a)}}}%
          }%
          \gplgaddtomacro\gplfronttext{%
            \csname LTb\endcsname%
            \put(63,2939){\rotatebox{-270}{\makebox(0,0){\strut{}Gap $\Delta(z)/\Delta_0$}}}%
            \csname LTb\endcsname%
            \put(1617,3921){\makebox(0,0){\strut{}P configuration}}%
            \colorrgb{0.00,0.00,0.00}%
            \put(666,2237){\makebox(0,0)[r]{\strut{}$\scalebox{0.8}{0.000}$}}%
            \colorrgb{0.00,0.00,0.00}%
            \put(666,2940){\makebox(0,0)[r]{\strut{}$\scalebox{0.8}{0.020}$}}%
            \colorrgb{0.00,0.00,0.00}%
            \put(666,3642){\makebox(0,0)[r]{\strut{}$\scalebox{0.8}{0.040}$}}%
            \colorrgb{0.00,0.00,0.00}%
            \put(768,2051){\makebox(0,0){\strut{}}}%
            \colorrgb{0.00,0.00,0.00}%
            \put(1617,2051){\makebox(0,0){\strut{}}}%
            \colorrgb{0.00,0.00,0.00}%
            \put(2466,2051){\makebox(0,0){\strut{}}}%
          }%
          \gplgaddtomacro\gplbacktext{%
            \colorrgb{0.00,0.00,0.00}%
            \put(2941,2237){\makebox(0,0)[r]{\strut{}$\scalebox{0.8}{0.000}$}}%
            \colorrgb{0.00,0.00,0.00}%
            \put(2941,2940){\makebox(0,0)[r]{\strut{}$\scalebox{0.8}{0.500}$}}%
            \colorrgb{0.00,0.00,0.00}%
            \put(2941,3642){\makebox(0,0)[r]{\strut{}$\scalebox{0.8}{1.000}$}}%
            \colorrgb{0.00,0.00,0.00}%
            \put(3043,2051){\makebox(0,0){\strut{}}}%
            \colorrgb{0.00,0.00,0.00}%
            \put(3892,2051){\makebox(0,0){\strut{}}}%
            \colorrgb{0.00,0.00,0.00}%
            \put(4741,2051){\makebox(0,0){\strut{}}}%
            \csname LTb\endcsname%
            \put(3094,3502){\makebox(0,0)[l]{\strut{}\textbf{(b)}}}%
          }%
          \gplgaddtomacro\gplfronttext{%
            \csname LTb\endcsname%
            \put(3892,3921){\makebox(0,0){\strut{}AP configuration}}%
            \colorrgb{0.00,0.00,0.00}%
            \put(2941,2237){\makebox(0,0)[r]{\strut{}$\scalebox{0.8}{0.000}$}}%
            \colorrgb{0.00,0.00,0.00}%
            \put(2941,2940){\makebox(0,0)[r]{\strut{}$\scalebox{0.8}{0.500}$}}%
            \colorrgb{0.00,0.00,0.00}%
            \put(2941,3642){\makebox(0,0)[r]{\strut{}$\scalebox{0.8}{1.000}$}}%
            \colorrgb{0.00,0.00,0.00}%
            \put(3043,2051){\makebox(0,0){\strut{}}}%
            \colorrgb{0.00,0.00,0.00}%
            \put(3892,2051){\makebox(0,0){\strut{}}}%
            \colorrgb{0.00,0.00,0.00}%
            \put(4741,2051){\makebox(0,0){\strut{}}}%
          }%
          \gplgaddtomacro\gplbacktext{%
            \colorrgb{0.00,0.00,0.00}%
            \put(666,621){\makebox(0,0)[r]{\strut{}$\scalebox{0.8}{-0.500}$}}%
            \colorrgb{0.00,0.00,0.00}%
            \put(666,1297){\makebox(0,0)[r]{\strut{}$\scalebox{0.8}{0.000}$}}%
            \colorrgb{0.00,0.00,0.00}%
            \put(666,1972){\makebox(0,0)[r]{\strut{}$\scalebox{0.8}{0.500}$}}%
            \colorrgb{0.00,0.00,0.00}%
            \put(768,408){\makebox(0,0){\strut{}$\scalebox{0.8}{-0.5}$}}%
            \colorrgb{0.00,0.00,0.00}%
            \put(1617,408){\makebox(0,0){\strut{}$\scalebox{0.8}{0.0}$}}%
            \colorrgb{0.00,0.00,0.00}%
            \put(2466,408){\makebox(0,0){\strut{}$\scalebox{0.8}{0.5}$}}%
            \csname LTb\endcsname%
            \put(819,1859){\makebox(0,0)[l]{\strut{}\textbf{(c)}}}%
          }%
          \gplgaddtomacro\gplfronttext{%
            \csname LTb\endcsname%
            \put(93,1296){\rotatebox{-270}{\makebox(0,0){\strut{}Phase $\varphi(z)/\pi$}}}%
            \csname LTb\endcsname%
            \put(1617,129){\makebox(0,0){\strut{}Position $z/L$}}%
            \colorrgb{0.00,0.00,0.00}%
            \put(666,621){\makebox(0,0)[r]{\strut{}$\scalebox{0.8}{-0.500}$}}%
            \colorrgb{0.00,0.00,0.00}%
            \put(666,1297){\makebox(0,0)[r]{\strut{}$\scalebox{0.8}{0.000}$}}%
            \colorrgb{0.00,0.00,0.00}%
            \put(666,1972){\makebox(0,0)[r]{\strut{}$\scalebox{0.8}{0.500}$}}%
            \colorrgb{0.00,0.00,0.00}%
            \put(768,408){\makebox(0,0){\strut{}$\scalebox{0.8}{-0.5}$}}%
            \colorrgb{0.00,0.00,0.00}%
            \put(1617,408){\makebox(0,0){\strut{}$\scalebox{0.8}{0.0}$}}%
            \colorrgb{0.00,0.00,0.00}%
            \put(2466,408){\makebox(0,0){\strut{}$\scalebox{0.8}{0.5}$}}%
          }%
          \gplgaddtomacro\gplbacktext{%
            \colorrgb{0.00,0.00,0.00}%
            \put(2941,594){\makebox(0,0)[r]{\strut{}$\scalebox{0.8}{-0.003}$}}%
            \colorrgb{0.00,0.00,0.00}%
            \put(2941,1297){\makebox(0,0)[r]{\strut{}$\scalebox{0.8}{0.000}$}}%
            \colorrgb{0.00,0.00,0.00}%
            \put(2941,1999){\makebox(0,0)[r]{\strut{}$\scalebox{0.8}{0.003}$}}%
            \colorrgb{0.00,0.00,0.00}%
            \put(3043,408){\makebox(0,0){\strut{}$\scalebox{0.8}{-0.5}$}}%
            \colorrgb{0.00,0.00,0.00}%
            \put(3892,408){\makebox(0,0){\strut{}$\scalebox{0.8}{0.0}$}}%
            \colorrgb{0.00,0.00,0.00}%
            \put(4741,408){\makebox(0,0){\strut{}$\scalebox{0.8}{0.5}$}}%
            \csname LTb\endcsname%
            \put(3094,1859){\makebox(0,0)[l]{\strut{}\textbf{(d)}}}%
          }%
          \gplgaddtomacro\gplfronttext{%
            \csname LTb\endcsname%
            \put(3892,129){\makebox(0,0){\strut{}Position $z/L$}}%
            \colorrgb{0.00,0.00,0.00}%
            \put(2941,594){\makebox(0,0)[r]{\strut{}$\scalebox{0.8}{-0.003}$}}%
            \colorrgb{0.00,0.00,0.00}%
            \put(2941,1297){\makebox(0,0)[r]{\strut{}$\scalebox{0.8}{0.000}$}}%
            \colorrgb{0.00,0.00,0.00}%
            \put(2941,1999){\makebox(0,0)[r]{\strut{}$\scalebox{0.8}{0.003}$}}%
            \colorrgb{0.00,0.00,0.00}%
            \put(3043,408){\makebox(0,0){\strut{}$\scalebox{0.8}{-0.5}$}}%
            \colorrgb{0.00,0.00,0.00}%
            \put(3892,408){\makebox(0,0){\strut{}$\scalebox{0.8}{0.0}$}}%
            \colorrgb{0.00,0.00,0.00}%
            \put(4741,408){\makebox(0,0){\strut{}$\scalebox{0.8}{0.5}$}}%
          }%
          \gplbacktext
          \put(0,0){\includegraphics{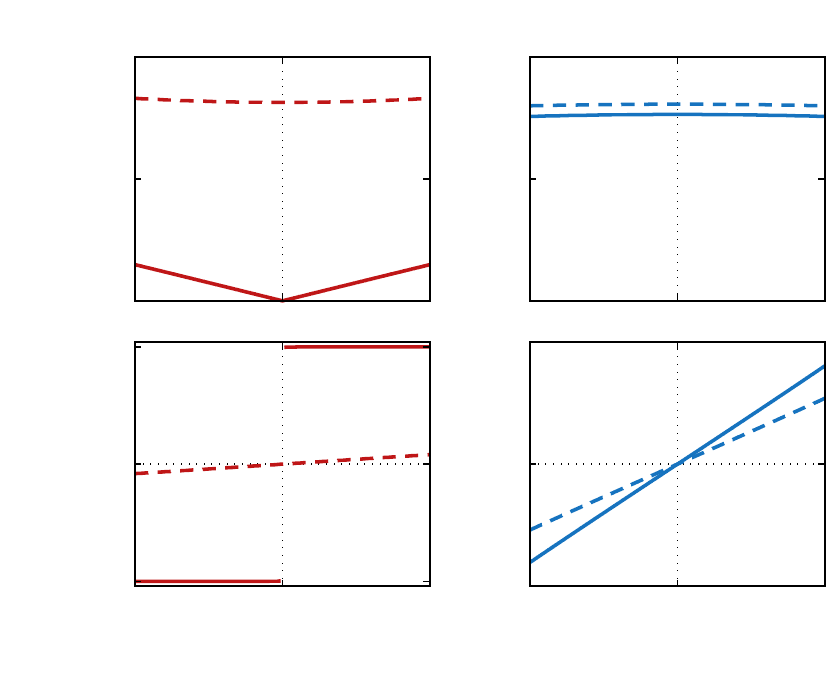}}%
          \gplfronttext
        \end{picture}%
      \endgroup
  \caption
  {%
    Numerical results for the superconducting order parameter $\Delta(z)\, e^{i\varphi(z)}${, where $\Delta$ and $\varphi$ are real-valued functions of the position~$z$ in the central superconductor.
    The dashed lines correspond to a phase-difference $\delta\varphi = \pi/2$, and the solid lines to $\delta\varphi = \pi$.
    In the AP case, the gap $\Delta(z)$ is close to the bulk gap $\Delta_0$ everywhere.
    We have a finite phase-winding for both $\pi/2$ and $\pi$ phase-difference, but it is larger in the latter case.
    In the P case, however, the gap is 1--2 orders of magnitude smaller, and even drops to zero at the center for $\delta\varphi = \pi$.
    Note the discontinuity in the phase for $\delta\varphi = \pi$ and $z=0$, which is where the order parameter $\Delta e^{i\varphi}$ changes sign.
    The gap plots are consistent with a spin-valve effect, since $\Delta$ is suppressed in the P but not AP configuration.
    The phase plots are consistent with \cref{fig:cpr}, since the current $J$ is proportional to the phase-winding~$\partial_z \varphi$ at every point.}
  }%
  \label{fig:gap}
\end{figure}

\noindent
We will start by considering a perfectly symmetric S/F/S/F/S structure, where the symmetry implies that the critical currents of the constituent S/F/S subjunctions are equal.
{In this subsection, we restrict our attention to the regime $-\pi < \delta\varphi < +\pi$, since the ground-state is known to be $2\pi$-periodic.}
The numerical results  are shown in \cref{fig:cpr,fig:gap}.
In \cref{fig:cpr}, we see that it is indeed possible to magnetically switch between a very clean $\sin(\delta\varphi)$ and $\sin(\delta\varphi/2)$ current-phase relation.
The transition between a conventional and fractional Josephson effect obtained in this manner is sensitive to the parameters of the junction.
In a regular spin-valve, achieving an absolute spin-valve effect in the junction requires that the superconductor length is short enough and the ferromagnetic properties strong enough to vanquish the superconducting condensate in the P configuration.
In the system under consideration here, an additional complication is that one simultaneously requires a sufficiently low conductance through the ferromagnetic insulators in order to be able to turn superconductivity off at all in the central superconductor.
Otherwise, the bulk superconductors are able to supply enough Cooper pairs via the proximity effect to keep the central layer superconducting regardless of the magnetic configuration. However, making the conductance too low would limit the supercurrents that we are interested in.
Therefore, in an experiment, some trial and error might be required to obtain ideal thicknesses for the material layers.

We proceed to discuss the physics underlying the transition between the conventional and fractional Josephson effect.
{From \cref{fig:gap}, we see that the gap is close to the bulk value $\Delta_0$ in the AP configuration, while it drops below $0.04\Delta_0$ in the P configuration.}
In other words, we have the desired spin-valve effect, where the magnetic configuration of the junction alone is enough to change the selfconsistently calculated gap in the central superconductor by orders of magnitude.
{Note that in all cases where a current is flowing, the phase-winding in the central superconductor is relatively small, which means that a large part of the phase-winding must be happening in the ferromagnetic insulators inbetween the superconductors.}
{In the P configuration with $\delta\varphi = \pi$, the order parameter $\Delta(z) e^{i\varphi(z)}$ changes linearly through the junction, and drops to zero at the center.
This is the same result as one would obtain for the proximity-induced minigap in a short $\pi$-biased S/N/S junction, meaning that this system indeed does act as such a junction.}

\subsection{Realistic case: asymmetric junction}\label{sec:num-sinphi2-asymm}
\begin{figure}[b!]
    \begingroup
    \makeatletter
    \providecommand\color[2][]{%
      \GenericError{(gnuplot) \space\space\space\@spaces}{%
        Package color not loaded in conjunction with
        terminal option `colourtext'%
      }{See the gnuplot documentation for explanation.%
      }{Either use 'blacktext' in gnuplot or load the package
        color.sty in LaTeX.}%
      \renewcommand\color[2][]{}%
    }%
    \providecommand\includegraphics[2][]{%
      \GenericError{(gnuplot) \space\space\space\@spaces}{%
        Package graphicx or graphics not loaded%
      }{See the gnuplot documentation for explanation.%
      }{The gnuplot epslatex terminal needs graphicx.sty or graphics.sty.}%
      \renewcommand\includegraphics[2][]{}%
    }%
    \providecommand\rotatebox[2]{#2}%
    \@ifundefined{ifGPcolor}{%
      \newif\ifGPcolor
      \GPcolortrue
    }{}%
    \@ifundefined{ifGPblacktext}{%
      \newif\ifGPblacktext
      \GPblacktexttrue
    }{}%
    \let\gplgaddtomacro\g@addto@macro
    \gdef\gplbacktext{}%
    \gdef\gplfronttext{}%
    \makeatother
    \ifGPblacktext
      \def\colorrgb#1{}%
      \def\colorgray#1{}%
    \else
      \ifGPcolor
        \def\colorrgb#1{\color[rgb]{#1}}%
        \def\colorgray#1{\color[gray]{#1}}%
        \expandafter\def\csname LTw\endcsname{\color{white}}%
        \expandafter\def\csname LTb\endcsname{\color{black}}%
        \expandafter\def\csname LTa\endcsname{\color{black}}%
        \expandafter\def\csname LT0\endcsname{\color[rgb]{1,0,0}}%
        \expandafter\def\csname LT1\endcsname{\color[rgb]{0,1,0}}%
        \expandafter\def\csname LT2\endcsname{\color[rgb]{0,0,1}}%
        \expandafter\def\csname LT3\endcsname{\color[rgb]{1,0,1}}%
        \expandafter\def\csname LT4\endcsname{\color[rgb]{0,1,1}}%
        \expandafter\def\csname LT5\endcsname{\color[rgb]{1,1,0}}%
        \expandafter\def\csname LT6\endcsname{\color[rgb]{0,0,0}}%
        \expandafter\def\csname LT7\endcsname{\color[rgb]{1,0.3,0}}%
        \expandafter\def\csname LT8\endcsname{\color[rgb]{0.5,0.5,0.5}}%
      \else
        \def\colorrgb#1{\color{black}}%
        \def\colorgray#1{\color[gray]{#1}}%
        \expandafter\def\csname LTw\endcsname{\color{white}}%
        \expandafter\def\csname LTb\endcsname{\color{black}}%
        \expandafter\def\csname LTa\endcsname{\color{black}}%
        \expandafter\def\csname LT0\endcsname{\color{black}}%
        \expandafter\def\csname LT1\endcsname{\color{black}}%
        \expandafter\def\csname LT2\endcsname{\color{black}}%
        \expandafter\def\csname LT3\endcsname{\color{black}}%
        \expandafter\def\csname LT4\endcsname{\color{black}}%
        \expandafter\def\csname LT5\endcsname{\color{black}}%
        \expandafter\def\csname LT6\endcsname{\color{black}}%
        \expandafter\def\csname LT7\endcsname{\color{black}}%
        \expandafter\def\csname LT8\endcsname{\color{black}}%
      \fi
    \fi
      \setlength{\unitlength}{0.0500bp}%
      \ifx\gptboxheight\undefined%
        \newlength{\gptboxheight}%
        \newlength{\gptboxwidth}%
        \newsavebox{\gptboxtext}%
      \fi%
      \setlength{\fboxrule}{0.5pt}%
      \setlength{\fboxsep}{1pt}%
    \begin{picture}(4800.00,1800.00)%
      \gplgaddtomacro\gplbacktext{%
        \colorrgb{0.00,0.00,0.00}%
        \put(378,600){\makebox(0,0)[r]{\strut{}$\scalebox{0.8}{-1}$}}%
        \colorrgb{0.00,0.00,0.00}%
        \put(378,1170){\makebox(0,0)[r]{\strut{}$\scalebox{0.8}{0}$}}%
        \colorrgb{0.00,0.00,0.00}%
        \put(378,1739){\makebox(0,0)[r]{\strut{}$\scalebox{0.8}{1}$}}%
        \colorrgb{0.00,0.00,0.00}%
        \put(480,408){\makebox(0,0){\strut{}$\scalebox{0.8}{-2}$}}%
        \colorrgb{0.00,0.00,0.00}%
        \put(914,408){\makebox(0,0){\strut{}$\scalebox{0.8}{-1}$}}%
        \colorrgb{0.00,0.00,0.00}%
        \put(1349,408){\makebox(0,0){\strut{}$\scalebox{0.8}{0}$}}%
        \colorrgb{0.00,0.00,0.00}%
        \put(1783,408){\makebox(0,0){\strut{}$\scalebox{0.8}{1}$}}%
        \colorrgb{0.00,0.00,0.00}%
        \put(2217,408){\makebox(0,0){\strut{}$\scalebox{0.8}{2}$}}%
        \csname LTb\endcsname%
        \put(506,1630){\makebox(0,0)[l]{\strut{}\textbf{(a)}}}%
      }%
      \gplgaddtomacro\gplfronttext{%
        \csname LTb\endcsname%
        \put(81,1169){\rotatebox{-270}{\makebox(0,0){\strut{}Current $J/J_c$}}}%
        \csname LTb\endcsname%
        \put(1348,129){\makebox(0,0){\strut{}Phase-difference $\delta\varphi/\pi$}}%
        \colorrgb{0.00,0.00,0.00}%
        \put(378,600){\makebox(0,0)[r]{\strut{}$\scalebox{0.8}{-1}$}}%
        \colorrgb{0.00,0.00,0.00}%
        \put(378,1170){\makebox(0,0)[r]{\strut{}$\scalebox{0.8}{0}$}}%
        \colorrgb{0.00,0.00,0.00}%
        \put(378,1739){\makebox(0,0)[r]{\strut{}$\scalebox{0.8}{1}$}}%
        \colorrgb{0.00,0.00,0.00}%
        \put(480,408){\makebox(0,0){\strut{}$\scalebox{0.8}{-2}$}}%
        \colorrgb{0.00,0.00,0.00}%
        \put(914,408){\makebox(0,0){\strut{}$\scalebox{0.8}{-1}$}}%
        \colorrgb{0.00,0.00,0.00}%
        \put(1349,408){\makebox(0,0){\strut{}$\scalebox{0.8}{0}$}}%
        \colorrgb{0.00,0.00,0.00}%
        \put(1783,408){\makebox(0,0){\strut{}$\scalebox{0.8}{1}$}}%
        \colorrgb{0.00,0.00,0.00}%
        \put(2217,408){\makebox(0,0){\strut{}$\scalebox{0.8}{2}$}}%
      }%
      \gplgaddtomacro\gplbacktext{%
        \colorrgb{0.00,0.00,0.00}%
        \put(2883,600){\makebox(0,0)[r]{\strut{}$\scalebox{0.8}{-1}$}}%
        \colorrgb{0.00,0.00,0.00}%
        \put(2883,1170){\makebox(0,0)[r]{\strut{}$\scalebox{0.8}{0}$}}%
        \colorrgb{0.00,0.00,0.00}%
        \put(2883,1739){\makebox(0,0)[r]{\strut{}$\scalebox{0.8}{1}$}}%
        \colorrgb{0.00,0.00,0.00}%
        \put(2985,408){\makebox(0,0){\strut{}$\scalebox{0.8}{-2}$}}%
        \colorrgb{0.00,0.00,0.00}%
        \put(3419,408){\makebox(0,0){\strut{}$\scalebox{0.8}{-1}$}}%
        \colorrgb{0.00,0.00,0.00}%
        \put(3854,408){\makebox(0,0){\strut{}$\scalebox{0.8}{0}$}}%
        \colorrgb{0.00,0.00,0.00}%
        \put(4288,408){\makebox(0,0){\strut{}$\scalebox{0.8}{1}$}}%
        \colorrgb{0.00,0.00,0.00}%
        \put(4722,408){\makebox(0,0){\strut{}$\scalebox{0.8}{2}$}}%
        \csname LTb\endcsname%
        \put(3015,1607){\makebox(0,0)[l]{\strut{}\textbf{(b)}}}%
      }%
      \gplgaddtomacro\gplfronttext{%
        \csname LTb\endcsname%
        \put(2596,1169){\rotatebox{-270}{\makebox(0,0){\strut{}Phase $\varphi_2/\pi$}}}%
        \csname LTb\endcsname%
        \put(3853,129){\makebox(0,0){\strut{}Phase-difference $\delta\varphi/\pi$}}%
        \colorrgb{0.00,0.00,0.00}%
        \put(2883,600){\makebox(0,0)[r]{\strut{}$\scalebox{0.8}{-1}$}}%
        \colorrgb{0.00,0.00,0.00}%
        \put(2883,1170){\makebox(0,0)[r]{\strut{}$\scalebox{0.8}{0}$}}%
        \colorrgb{0.00,0.00,0.00}%
        \put(2883,1739){\makebox(0,0)[r]{\strut{}$\scalebox{0.8}{1}$}}%
        \colorrgb{0.00,0.00,0.00}%
        \put(2985,408){\makebox(0,0){\strut{}$\scalebox{0.8}{-2}$}}%
        \colorrgb{0.00,0.00,0.00}%
        \put(3419,408){\makebox(0,0){\strut{}$\scalebox{0.8}{-1}$}}%
        \colorrgb{0.00,0.00,0.00}%
        \put(3854,408){\makebox(0,0){\strut{}$\scalebox{0.8}{0}$}}%
        \colorrgb{0.00,0.00,0.00}%
        \put(4288,408){\makebox(0,0){\strut{}$\scalebox{0.8}{1}$}}%
        \colorrgb{0.00,0.00,0.00}%
        \put(4722,408){\makebox(0,0){\strut{}$\scalebox{0.8}{2}$}}%
      }%
      \gplbacktext
      \put(0,0){\includegraphics{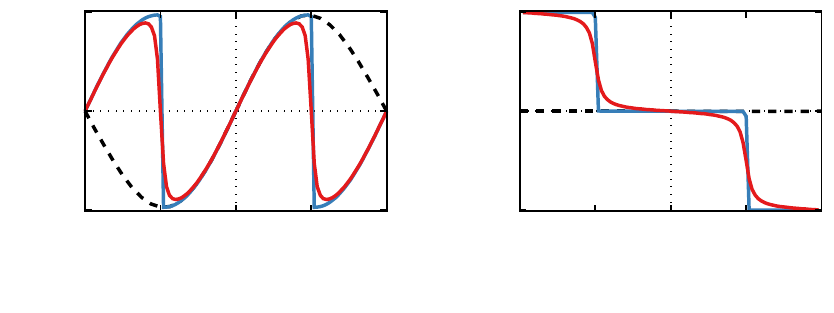}}%
      \gplfronttext
    \end{picture}%
    \endgroup
  \caption
	{
		Numerical results for \textbf{(a)} the current-phase relation and \textbf{(b)} the central phase $\varphi_2$ as functions of the external phase-difference~$\delta\varphi$.
		Both plots are for asymmetric junctions with interface conductances $G_T=0.02G_0(1 \pm a)$, with $a$ set to 0 (dashed), 0.0005 (blue), 0.1 (red).\\
    {The physical current-phase relation in the ground state is always $2\pi$-periodic due to discontinuities at $\delta\varphi=\pm\pi$, but asymmetries in realistic junctions smooth these discontinuous transitions.}
	}
	\label{fig:asymmetry}
\end{figure}

\noindent
It is impossible to construct a perfectly symmetric S/X/S/X/S junction {in reality}, and it is therefore prudent to check how sensitive the proposed setup is to asymmetries.
We have done so by setting the tunneling conductance to $G_T=0.02G_0(1 \pm a)$ at the left and right interface, respectively.
Here, $a$ parametrizes the asymmetry in the junction.
It is easy to see that in the limit $a\rightarrow0$ we regain the symmetric case, while $a\rightarrow1$ would decouple the central superconductor entirely from one of the electrodes.
{As we see from the results in \cref{fig:asymmetry}, the main effect of the asymmetry is to soften the discontinuities at $\delta\varphi=\pm\pi$.}

{For the perfectly symmetric case $a=0$, we obtained a $4\pi$-periodic result, since the numerics converged to the excited branch $J = -J_c \sin(\delta\varphi/2) \mathrm{sgn}[\cos(\delta\varphi/2)]$ instead of the ground state $J = +J_c \sin(\delta\varphi/2) \mathrm{sgn}[\cos(\delta\varphi/2)]$ for $|\delta\varphi|>\pi$.}
{However, this excited state is unstable, and introducing even a tiny asymmetry $a=0.0005$ results in the numerics correctly converging towards the $2\pi$-periodic ground state.
This ground state is characterized by abrupt discontinuities for $\delta\varphi=\pm\pi$ [\cref{fig:asymmetry}(a)], accompanied by equally abrupt changes in the phase of the central superconductor from $\varphi_2=0$ to $\varphi_2=\pm\pi$ [\cref{fig:asymmetry}(b)].}
As the asymmetry~$a$ increases, the discontinuities become smoother.

If we take the superconductors in the junction to be Nb, then the density of states at the Fermi level $N_0 \approx 80$~eV$^{-1}$\,nm$^{-3}$,\cite{Jani1988} the zero-temperature gap $\Delta_0 \approx 1.4$~meV, and a typical value for the coherence length $\xi \approx 15$~nm.
We assumed that the junction is at a temperature much lower than the critical temperature $T/T_c=0.01$, and has an inelastic scattering rate $\delta/\Delta_0 = 0.01$.
The Debye cutoff was set to $\Theta = 30\Delta_0$, which is high enough for the results to be independent of the cutoff.
Furthermore, we assumed that the length of the central superconductor was $L = \xi/2 \approx 7.5$~nm.
Putting these values together, and dividing by the cross-sectional area $A$ of the junction, we find that the current density unit $J_0/A \approx 3 \times 10^{7}$~A/cm$^2$.
Since we in \cref{fig:cpr} found critical currents between $J_0/4$~(AP) and $J_0/200$~(P), this translates to current densities of $\sim\! 10^{7}$~A/cm$^2$ and $\sim\! 10^{5}$~A/cm$^2$, respectively.
Note that this likely overestimates the current densities one would observe experimentally, since we treat the superconductors at the end of the junction as bulk superconductors.
In reality, one might expect the order parameter to be suppressed near the interface to a strongly polarized magnetic insulator~\cite{Ouassou2016b}, which would throttle the current.

\subsection{Application: spin-switch junction}\noindent
\begin{figure}[t]
	\hspace{-3.6em}
      \begingroup
      \makeatletter
      \providecommand\color[2][]{%
        \GenericError{(gnuplot) \space\space\space\@spaces}{%
          Package color not loaded in conjunction with
          terminal option `colourtext'%
        }{See the gnuplot documentation for explanation.%
        }{Either use 'blacktext' in gnuplot or load the package
          color.sty in LaTeX.}%
        \renewcommand\color[2][]{}%
      }%
      \providecommand\includegraphics[2][]{%
        \GenericError{(gnuplot) \space\space\space\@spaces}{%
          Package graphicx or graphics not loaded%
        }{See the gnuplot documentation for explanation.%
        }{The gnuplot epslatex terminal needs graphicx.sty or graphics.sty.}%
        \renewcommand\includegraphics[2][]{}%
      }%
      \providecommand\rotatebox[2]{#2}%
      \@ifundefined{ifGPcolor}{%
        \newif\ifGPcolor
        \GPcolortrue
      }{}%
      \@ifundefined{ifGPblacktext}{%
        \newif\ifGPblacktext
        \GPblacktexttrue
      }{}%
      \let\gplgaddtomacro\g@addto@macro
      \gdef\gplbacktext{}%
      \gdef\gplfronttext{}%
      \makeatother
      \ifGPblacktext
        \def\colorrgb#1{}%
        \def\colorgray#1{}%
      \else
        \ifGPcolor
          \def\colorrgb#1{\color[rgb]{#1}}%
          \def\colorgray#1{\color[gray]{#1}}%
          \expandafter\def\csname LTw\endcsname{\color{white}}%
          \expandafter\def\csname LTb\endcsname{\color{black}}%
          \expandafter\def\csname LTa\endcsname{\color{black}}%
          \expandafter\def\csname LT0\endcsname{\color[rgb]{1,0,0}}%
          \expandafter\def\csname LT1\endcsname{\color[rgb]{0,1,0}}%
          \expandafter\def\csname LT2\endcsname{\color[rgb]{0,0,1}}%
          \expandafter\def\csname LT3\endcsname{\color[rgb]{1,0,1}}%
          \expandafter\def\csname LT4\endcsname{\color[rgb]{0,1,1}}%
          \expandafter\def\csname LT5\endcsname{\color[rgb]{1,1,0}}%
          \expandafter\def\csname LT6\endcsname{\color[rgb]{0,0,0}}%
          \expandafter\def\csname LT7\endcsname{\color[rgb]{1,0.3,0}}%
          \expandafter\def\csname LT8\endcsname{\color[rgb]{0.5,0.5,0.5}}%
        \else
          \def\colorrgb#1{\color{black}}%
          \def\colorgray#1{\color[gray]{#1}}%
          \expandafter\def\csname LTw\endcsname{\color{white}}%
          \expandafter\def\csname LTb\endcsname{\color{black}}%
          \expandafter\def\csname LTa\endcsname{\color{black}}%
          \expandafter\def\csname LT0\endcsname{\color{black}}%
          \expandafter\def\csname LT1\endcsname{\color{black}}%
          \expandafter\def\csname LT2\endcsname{\color{black}}%
          \expandafter\def\csname LT3\endcsname{\color{black}}%
          \expandafter\def\csname LT4\endcsname{\color{black}}%
          \expandafter\def\csname LT5\endcsname{\color{black}}%
          \expandafter\def\csname LT6\endcsname{\color{black}}%
          \expandafter\def\csname LT7\endcsname{\color{black}}%
          \expandafter\def\csname LT8\endcsname{\color{black}}%
        \fi
      \fi
        \setlength{\unitlength}{0.0500bp}%
        \ifx\gptboxheight\undefined%
          \newlength{\gptboxheight}%
          \newlength{\gptboxwidth}%
          \newsavebox{\gptboxtext}%
        \fi%
        \setlength{\fboxrule}{0.5pt}%
        \setlength{\fboxsep}{1pt}%
      \begin{picture}(3760.00,6800.00)%
        \gplgaddtomacro\gplbacktext{%
          \colorrgb{0.00,0.00,0.00}%
          \put(969,5150){\makebox(0,0)[r]{\strut{}$\scalebox{0.8}{1}$}}%
          \colorrgb{0.00,0.00,0.00}%
          \put(969,5625){\makebox(0,0)[r]{\strut{}$\scalebox{0.8}{10}$}}%
          \colorrgb{0.00,0.00,0.00}%
          \put(969,6100){\makebox(0,0)[r]{\strut{}$\scalebox{0.8}{100}$}}%
          \colorrgb{0.00,0.00,0.00}%
          \put(969,6575){\makebox(0,0)[r]{\strut{}$\scalebox{0.8}{1000}$}}%
          \colorrgb{0.00,0.00,0.00}%
          \put(1071,4942){\makebox(0,0){\strut{}$\scalebox{0.8}{0}$}}%
          \colorrgb{0.00,0.00,0.00}%
          \put(1547,4942){\makebox(0,0){\strut{}$\scalebox{0.8}{0.1}$}}%
          \colorrgb{0.00,0.00,0.00}%
          \put(2024,4942){\makebox(0,0){\strut{}$\scalebox{0.8}{0.2}$}}%
          \colorrgb{0.00,0.00,0.00}%
          \put(2500,4942){\makebox(0,0){\strut{}$\scalebox{0.8}{0.3}$}}%
          \colorrgb{0.00,0.00,0.00}%
          \put(2977,4942){\makebox(0,0){\strut{}$\scalebox{0.8}{0.4}$}}%
          \colorrgb{0.00,0.00,0.00}%
          \put(3453,4942){\makebox(0,0){\strut{}$\scalebox{0.8}{0.5}$}}%
          \csname LTb\endcsname%
          \put(1142,6448){\makebox(0,0)[l]{\strut{}\textbf{(a)}}}%
        }%
        \gplgaddtomacro\gplfronttext{%
          \csname LTb\endcsname%
          \put(570,5861){\rotatebox{-270}{\makebox(0,0){\strut{}On--off ratio}}}%
          \csname LTb\endcsname%
          \put(2262,4663){\makebox(0,0){\strut{}Tunneling conductance $G_T/G_0$}}%
          \colorrgb{0.00,0.00,0.00}%
          \put(969,5150){\makebox(0,0)[r]{\strut{}$\scalebox{0.8}{1}$}}%
          \colorrgb{0.00,0.00,0.00}%
          \put(969,5625){\makebox(0,0)[r]{\strut{}$\scalebox{0.8}{10}$}}%
          \colorrgb{0.00,0.00,0.00}%
          \put(969,6100){\makebox(0,0)[r]{\strut{}$\scalebox{0.8}{100}$}}%
          \colorrgb{0.00,0.00,0.00}%
          \put(969,6575){\makebox(0,0)[r]{\strut{}$\scalebox{0.8}{1000}$}}%
          \colorrgb{0.00,0.00,0.00}%
          \put(1071,4942){\makebox(0,0){\strut{}$\scalebox{0.8}{0}$}}%
          \colorrgb{0.00,0.00,0.00}%
          \put(1547,4942){\makebox(0,0){\strut{}$\scalebox{0.8}{0.1}$}}%
          \colorrgb{0.00,0.00,0.00}%
          \put(2024,4942){\makebox(0,0){\strut{}$\scalebox{0.8}{0.2}$}}%
          \colorrgb{0.00,0.00,0.00}%
          \put(2500,4942){\makebox(0,0){\strut{}$\scalebox{0.8}{0.3}$}}%
          \colorrgb{0.00,0.00,0.00}%
          \put(2977,4942){\makebox(0,0){\strut{}$\scalebox{0.8}{0.4}$}}%
          \colorrgb{0.00,0.00,0.00}%
          \put(3453,4942){\makebox(0,0){\strut{}$\scalebox{0.8}{0.5}$}}%
        }%
        \gplgaddtomacro\gplbacktext{%
          \colorrgb{0.00,0.00,0.00}%
          \put(969,2883){\makebox(0,0)[r]{\strut{}$\scalebox{0.8}{1}$}}%
          \colorrgb{0.00,0.00,0.00}%
          \put(969,3358){\makebox(0,0)[r]{\strut{}$\scalebox{0.8}{10}$}}%
          \colorrgb{0.00,0.00,0.00}%
          \put(969,3834){\makebox(0,0)[r]{\strut{}$\scalebox{0.8}{100}$}}%
          \colorrgb{0.00,0.00,0.00}%
          \put(969,4309){\makebox(0,0)[r]{\strut{}$\scalebox{0.8}{1000}$}}%
          \colorrgb{0.00,0.00,0.00}%
          \put(1071,2675){\makebox(0,0){\strut{}$\scalebox{0.8}{0.1}$}}%
          \colorrgb{0.00,0.00,0.00}%
          \put(1667,2675){\makebox(0,0){\strut{}$\scalebox{0.8}{0.2}$}}%
          \colorrgb{0.00,0.00,0.00}%
          \put(2262,2675){\makebox(0,0){\strut{}$\scalebox{0.8}{0.3}$}}%
          \colorrgb{0.00,0.00,0.00}%
          \put(2858,2675){\makebox(0,0){\strut{}$\scalebox{0.8}{0.4}$}}%
          \colorrgb{0.00,0.00,0.00}%
          \put(3453,2675){\makebox(0,0){\strut{}$\scalebox{0.8}{0.5}$}}%
          \csname LTb\endcsname%
          \put(1131,4182){\makebox(0,0)[l]{\strut{}\textbf{(b)}}}%
        }%
        \gplgaddtomacro\gplfronttext{%
          \csname LTb\endcsname%
          \put(570,3595){\rotatebox{-270}{\makebox(0,0){\strut{}On--off ratio}}}%
          \csname LTb\endcsname%
          \put(2262,2396){\makebox(0,0){\strut{}Spin-mixing conductance $G_\varphi/G_0$}}%
          \colorrgb{0.00,0.00,0.00}%
          \put(969,2883){\makebox(0,0)[r]{\strut{}$\scalebox{0.8}{1}$}}%
          \colorrgb{0.00,0.00,0.00}%
          \put(969,3358){\makebox(0,0)[r]{\strut{}$\scalebox{0.8}{10}$}}%
          \colorrgb{0.00,0.00,0.00}%
          \put(969,3834){\makebox(0,0)[r]{\strut{}$\scalebox{0.8}{100}$}}%
          \colorrgb{0.00,0.00,0.00}%
          \put(969,4309){\makebox(0,0)[r]{\strut{}$\scalebox{0.8}{1000}$}}%
          \colorrgb{0.00,0.00,0.00}%
          \put(1071,2675){\makebox(0,0){\strut{}$\scalebox{0.8}{0.1}$}}%
          \colorrgb{0.00,0.00,0.00}%
          \put(1667,2675){\makebox(0,0){\strut{}$\scalebox{0.8}{0.2}$}}%
          \colorrgb{0.00,0.00,0.00}%
          \put(2262,2675){\makebox(0,0){\strut{}$\scalebox{0.8}{0.3}$}}%
          \colorrgb{0.00,0.00,0.00}%
          \put(2858,2675){\makebox(0,0){\strut{}$\scalebox{0.8}{0.4}$}}%
          \colorrgb{0.00,0.00,0.00}%
          \put(3453,2675){\makebox(0,0){\strut{}$\scalebox{0.8}{0.5}$}}%
        }%
        \gplgaddtomacro\gplbacktext{%
          \colorrgb{0.00,0.00,0.00}%
          \put(969,617){\makebox(0,0)[r]{\strut{}$\scalebox{0.8}{1}$}}%
          \colorrgb{0.00,0.00,0.00}%
          \put(969,1092){\makebox(0,0)[r]{\strut{}$\scalebox{0.8}{10}$}}%
          \colorrgb{0.00,0.00,0.00}%
          \put(969,1567){\makebox(0,0)[r]{\strut{}$\scalebox{0.8}{100}$}}%
          \colorrgb{0.00,0.00,0.00}%
          \put(969,2042){\makebox(0,0)[r]{\strut{}$\scalebox{0.8}{1000}$}}%
          \colorrgb{0.00,0.00,0.00}%
          \put(1071,409){\makebox(0,0){\strut{}$\scalebox{0.8}{0.3}$}}%
          \colorrgb{0.00,0.00,0.00}%
          \put(1547,409){\makebox(0,0){\strut{}$\scalebox{0.8}{0.4}$}}%
          \colorrgb{0.00,0.00,0.00}%
          \put(2024,409){\makebox(0,0){\strut{}$\scalebox{0.8}{0.5}$}}%
          \colorrgb{0.00,0.00,0.00}%
          \put(2500,409){\makebox(0,0){\strut{}$\scalebox{0.8}{0.6}$}}%
          \colorrgb{0.00,0.00,0.00}%
          \put(2977,409){\makebox(0,0){\strut{}$\scalebox{0.8}{0.7}$}}%
          \colorrgb{0.00,0.00,0.00}%
          \put(3453,409){\makebox(0,0){\strut{}$\scalebox{0.8}{0.8}$}}%
          \csname LTb\endcsname%
          \put(1119,1915){\makebox(0,0)[l]{\strut{}\textbf{(c)}}}%
        }%
        \gplgaddtomacro\gplfronttext{%
          \csname LTb\endcsname%
          \put(570,1328){\rotatebox{-270}{\makebox(0,0){\strut{}On--off ratio}}}%
          \csname LTb\endcsname%
          \put(2262,130){\makebox(0,0){\strut{}Superconductor length $L/\xi$}}%
          \colorrgb{0.00,0.00,0.00}%
          \put(969,617){\makebox(0,0)[r]{\strut{}$\scalebox{0.8}{1}$}}%
          \colorrgb{0.00,0.00,0.00}%
          \put(969,1092){\makebox(0,0)[r]{\strut{}$\scalebox{0.8}{10}$}}%
          \colorrgb{0.00,0.00,0.00}%
          \put(969,1567){\makebox(0,0)[r]{\strut{}$\scalebox{0.8}{100}$}}%
          \colorrgb{0.00,0.00,0.00}%
          \put(969,2042){\makebox(0,0)[r]{\strut{}$\scalebox{0.8}{1000}$}}%
          \colorrgb{0.00,0.00,0.00}%
          \put(1071,409){\makebox(0,0){\strut{}$\scalebox{0.8}{0.3}$}}%
          \colorrgb{0.00,0.00,0.00}%
          \put(1547,409){\makebox(0,0){\strut{}$\scalebox{0.8}{0.4}$}}%
          \colorrgb{0.00,0.00,0.00}%
          \put(2024,409){\makebox(0,0){\strut{}$\scalebox{0.8}{0.5}$}}%
          \colorrgb{0.00,0.00,0.00}%
          \put(2500,409){\makebox(0,0){\strut{}$\scalebox{0.8}{0.6}$}}%
          \colorrgb{0.00,0.00,0.00}%
          \put(2977,409){\makebox(0,0){\strut{}$\scalebox{0.8}{0.7}$}}%
          \colorrgb{0.00,0.00,0.00}%
          \put(3453,409){\makebox(0,0){\strut{}$\scalebox{0.8}{0.8}$}}%
        }%
        \gplbacktext
        \put(0,0){\includegraphics{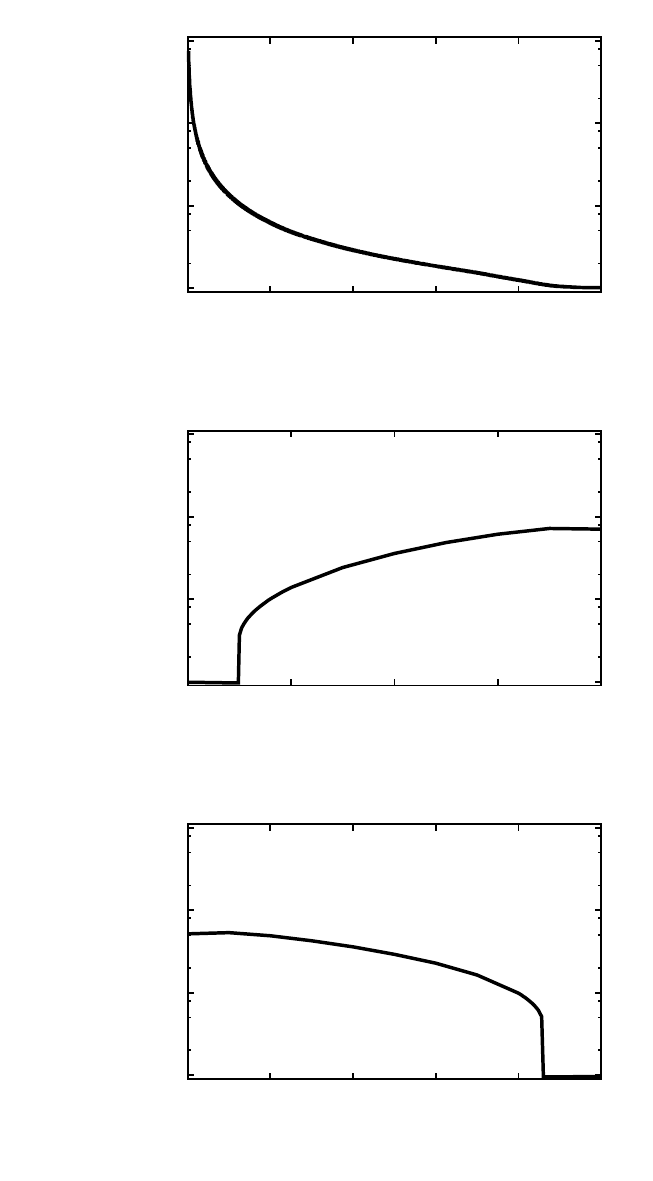}}%
        \gplfronttext
      \end{picture}%
      \endgroup
  \caption
	{
			Numerical results for the charge current on--off ratio $J^{\text{AP}}/J^{\text{P}}$ as functions of \textbf{(a)} the tunneling conductance~$G_T$, \textbf{(b)} the spin-mixing conductance~$G_\varphi$, and \textbf{(c)} the length of the central superconductor~$L$.
	}
	\label{fig:on-off}
\end{figure}
A possible application of the proposed device, which works well in both the ideal and realistic regimes discussed in the previous sections, is as a junction with a magnetically controlled on--off switch for the supercurrent.
\Cref{fig:cpr} shows that using the interface conductance $G_T=0.02G_0$, spin-mixing conductance $G_\varphi = 0.3G_0$, and superconductor length $L = \xi/2$, one can already obtain a factor $\sim\!\!50$ difference between the critical current in the P and AP configurations.
To further investigate this prospect, we {varied} one of these parameters at a time while keeping the others fixed.
For each parameter set, we calculated the ratio $J^{\text{AP}}/J^{\text{P}}$ between the charge current in the AP and P configurations, and the results are shown in \cref{fig:on-off}.
These calculations were done for a phase-difference $\delta\varphi=\pi/2$, which means that $J^{\text{AP}}/J^{\text{P}}$ is not strictly equal to the critical current ratio $J^{\text{AP}}_c/J^{\text{P}}_c$, but they should be the same order of magnitude.

In \cref{fig:on-off}(a), we see that for large interface conductances, the current ratio tends to one; and conversely, for very small interface conductances, the ratio diverges.
Thus, with respect to the on--off ratio, low interface conductances are the ideal choice.
On the other hand, a low interface conductance also means that the current is throttled even when the junction is turned on.
In practice, there will therefore be a trade-off between having a large on--off ratio (small conductance) and having a large current magnitude in the on-state (large conductance).

In \cref{fig:on-off}(b), we see that for small spin-mixing conductances~$G_\varphi$, the on--off ratio tends to one.
This is because in this limit, the spin-active properties are too weak to suppress superconductivity in either the P or AP states, leading to a large supercurrent in both the P and AP configurations.
After $G_\varphi \approx 0.15G_0$, the gap in the central superconductor is strongly suppressed in the P configuration but not the AP configuration, resulting in very high on--off ratios.
For very high spin-mixing conductances $G_\varphi \approx 0.9G_0$ (not shown), we also found a ${0\text{--}\pi}$ transition with a $\sin(2\delta\varphi)$ current-phase relation in the AP configuration, resulting in an even higher on--off ratio.

In \cref{fig:on-off}(c), we see that for superconductors larger than about $0.7\xi$, the on--off ratio tends to one.
This is because in this limit, the superconductor is large enough to sustain a significant gap in both the P and AP configurations, enabling large supercurrents to in both configurations.
On the other hand, for thin superconductors, the gap is strongly suppressed in the P but not AP configuration, yielding high on--off ratios.

\newpage
\section{Discussion}\noindent
The results in \cref{fig:cpr} confirm that it should be possible to toggle between {current-phase relations with} clear $\sin(\delta\varphi)${-} and $\sin(\delta\varphi/2)${-shapes} magnetically using experimentally realistic parameters.
{The current-phase relation remains $2\pi$-periodic in the ground state due to discontinuities at $\delta\varphi=\pm\pi$, which are smoothed out as the junctions become more asymmetric.}

Note that there is a factor 50 difference between the critical current in the P and AP states of \cref{fig:cpr}.
For potential applications where it is mainly the shape and not size of the current-phase relation that matters, this is of course not an obstacle.
For applications where the size of the supercurrent is important as well, the setup we propose may instead be considered as a $\sin(\delta\varphi/2)$ junction with a magnetic on--off switch, similar to the ideas in Refs.~\onlinecite{Birge2016,Baek2014,Bolginov2012}.
As shown in \cref{fig:on-off}, we find that this kind of setup can produce very high on--off ratios of 100--1000, as long as the superconductor is short enough and the spin-mixing high enough, in line with conventional wisdom regarding optimal spin-valve design.
The tunneling conductance was found to be the limiting ingredient, with lower conductances consistently resulting in higher on--off ratios.
However, at the same time, lower interface conductances mean lower currents in both the on and off states.
Thus, there is a tradeoff between acheiving high on--off ratios and high current densities in the on-state.

In principle, it might be possible to create a similar device using magnetic metals instead of insulators, since it has been experimentally demonstrated that superconducting spin-valves can be constructed out of metals too~\cite{gu_prl_02}.
However, it is then critical that the net tunneling conductance between each superconducting layer is small enough.
If the tunneling conductance is too high, then one would end up with both $\sin(\delta\varphi)${-} and $\sin(\delta\varphi/2)${-shaped} contributions in the AP configuration instead of a pure $\sin(\delta\varphi/2)${-shaped current-phase} relation.

Although we have focused on using a magnetic field to toggle between a $\sin(\delta\varphi)${-} and $\sin(\delta\varphi/2)${-shaped} current-phase relation so far, the same basic idea can be extended to other physical setups as well.
For instance, consider an S/I/S'/I/S structure, where S is a superconductor with a particular critical temperature (\eg Nb with $T_c \approx 9.2$~K), S' a superconductor with a lower critical temperature (\eg Al with $T_c' \approx 1.2$~K), and the I are thin layers of nonmagnetic insulators.
This junction should act as an S/X/S system with a $\sin(\delta\varphi)$ relation above $T_c'$, but as an S/X/S/X/S system with a $\sin(\delta\varphi/2)$ relation below $T_c'$ when the interface conductance is low enough to permit most of the phase-winding to occur at the interfaces {[as in \cref{fig:gap}]}.
In other words, it should also be possible to thermally toggle between these $\sin(\delta\varphi/n)${-shaped} current-phase relation{s}.\cite{Kupriyanov1999}\\


\newpage
\section{Conclusion}\noindent
We have demonstrated theoretically that by changing the magnetic configuration from a P to AP alignment in a spin-valve Josephson junction (S/F/S/F/S), it is possible to toggle between a $\sin(\delta\varphi)${-} and $\sin(\delta\varphi/2)${-shaped} current-phase {relation,} which retains a $2\pi$-periodicity {due to discontinuities} at $\delta\varphi=\pm\pi$.
The same mechanism may be used to effectively construct a Josephson junction with an on--off switch, acheiving 2--3 orders of magnitude difference between the on and off states.
We have also argued that the same procedure can be generalized to create arbitrary $\sin(\delta\varphi/n)$ junctions where $n$ is a magnetically tunable integer.
This is a novel way to exert control over superconductivity in nanoscale structures, which may both spur new fundamental research in superconducting spintronics,\cite{linder_nphys_15} and find practical applications in future cryogenic technology.
\newpage

\begin{acknowledgments}
  \noindent
  We wish to thank M. Amundsen and M. Yu. Kupriyanov for useful discussions.
  J.L. and J.A.O. were supported by the Research Council of Norway (grant 240806).
  J.L. was also supported by the Research Council of Norway (grant 216700), and the \emph{Outstanding Academic Fellows} programme at NTNU.
  The Research Council of Norway and NTNU are acknowledged for funding the Center of Excellence \textit{QuSpin}.
\end{acknowledgments}


\end{document}